\documentclass[a4paper]{scrartcl}
\usepackage[utf8]{inputenc}
\usepackage{natbib}
\usepackage{amsthm, amssymb, amsmath}
\usepackage{bm}
\usepackage{dsfont}
\usepackage{enumerate}
\usepackage[usenames,dvipsnames]{color}
\usepackage{subfigure}
\usepackage{caption}
\usepackage{graphicx}
\usepackage[unicode,breaklinks]{hyperref}

\definecolor{linkcolor}{RGB}{15,15,175}
\definecolor{alarmcolor}{RGB}{175,15,15}

\hypersetup{
    pdfauthor={Christian Sch\"afer},
    pdfcreator={Christian Sch\"afer},
    unicode=true,           
    pdftoolbar=true,        
    pdfmenubar=true,        
    pdffitwindow=false,     
    pdfstartview={FitH},    
    pdfnewwindow=true,      
    colorlinks=true,        
    linkcolor=linkcolor,    
    citecolor=linkcolor,    
    filecolor=linkcolor,    
    urlcolor =linkcolor     
}

\def\B{\mathbb{B}}

\def\R{\mathbb{R}}

\def\N{\mathbb{N}}
\def\Z{\mathbb{Z}}

\def\uni{\mathcal{U}} 
\def\cau{\mathcal{C}} 

\def\diag{\mathrm{diag}}
\def\tr#1{\mathrm{tr}\left( #1 \right)}
\def\abs#1{\left\vert #1 \right\vert}
\def\det{\mathrm{det}}
\def\set#1{\lbrace #1 \rbrace}
\def\dset#1{[\![#1]\!]}
\def\norm#1{\Vert #1 \Vert}
\def\card#1{\vert #1 \vert}
\def\t{^\intercal}
\def\argmax{\mathop{\mathrm{argmax}}}

\def\v#1{\bm{#1}}
\def\m#1{\bm{\mathrm{#1}}}

\def\eqdef{:=}

\def\ind{\mathds{1}}
\def\logistic{\ell}
\def\ls{L^{+}}

\newcommand{\ev}[1]{\mathbb{E}\left(#1 \right)}

\makeatletter
\newcommand{\raisemath}[1]{\mathpalette{\raisem@th{#1}}}
\newcommand{\raisem@th}[3]{\raisebox{#1}[0pt][0pt]{$#2#3$}}
\makeatother

\def\Prod{{\raisemath{-1.1pt}{\sqcap}}}

\def\LogCo{{\raisemath{-1.1pt}{\,\logistic}}}
\def\Gau{{\raisemath{-1.1pt}{gc}}}

\newenvironment{definition}[1][Definition]{\begin{trivlist}
\item[\hskip \labelsep {\bfseries #1}]}{\end{trivlist}}

\def\keywords#1{\par\addvspace\medskipamount{\rightskip=0pt plus1cm
\def\and{\ifhmode\unskip\nobreak\fi\ $\cdot$
}\noindent \textbf{Keywords}\enspace\ignorespaces#1\par}}

\addtokomafont{footnote}{\sffamily\fontsize{6}{3}\selectfont}
\author{Christian Sch\"afer$^{1,2}$}
\def\crest{
\footnotetext[1]{Centre de Recherche en Économie et Statistique, 3 Avenue Pierre Larousse, 92240 Malakoff, France}
\footnotetext[2]{CEntre de REcherches en MAthématiques de la DEcision, Université Paris-Dauphine, Place du Maréchal de Lattre de Tassigny
75775 Paris, France}
}

\usepackage[ruled,algo2e]{algorithm2e}
\renewcommand{\algorithmcfname}{Procedure}
\SetAlFnt{\small}
\SetAlCapFnt{\small}
\SetAlCapNameFnt{\small}
\SetAlCapHSkip{0pt}
\newenvironment{algorithm}{
\begin{figure}[ht]\begin{center}\begin{minipage}{0.75\textwidth}\begin{algorithm2e}[H]}
{\end{algorithm2e}\end{minipage}\end{center}\end{figure}}

\newcommand{\mygraph}[2]{
\subfigure[#1 problem #2]{
\includegraphics[width=0.499\textwidth]{#1_#2}
\hspace{-5mm}
}}%

\def\citeN#1{\citet{#1}}
\def\citeNP#1{\citep{#1}}

\title{Particle algorithms for optimization on binary spaces}

\hypersetup{
    pdftitle={Particle algorithms for optimization on binary spaces},
    pdfkeywords={Binary parametric families}{Unconstrained binary optimization}{Sequential Monte Carlo}{Cross-Entropy method}{Simulated Annealing}
}

\begin{document}

\maketitle
\crest
\thispagestyle{empty}

\begin{abstract}
We discuss a unified approach to stochastic optimization of pseudo-Boolean objective functions based on particle methods, including the cross-entropy method and simulated annealing as special cases. We point out the need for auxiliary sampling distributions, that is parametric families on binary spaces, which are able to reproduce complex dependency structures, and illustrate their usefulness in our numerical experiments. We provide numerical evidence that particle-driven optimization algorithms based on parametric families yield superior results on strongly multi-modal optimization problems while local search heuristics outperform them on easier problems.
\end{abstract}


\section{Particle optimization}
\subsection{Introduction}
\subsubsection{Pseudo-Boolean optimization}
We call $f\colon\B^d\eqdef\set{0,1}^d\to\R$ a pseudo-Boolean function. The present work discusses approaches to obtain heuristics for the program
\begin{equation}
\label{eq:pb program}
\begin{tabular}{ll}
\text{maximize }  & $f(\v x)$ \\[0.2em]
\text{subject to} & $ \v x\in\B^d$
\end{tabular}
\end{equation}
using sequential Monte Carlo techniques. In the sequel, we refer to $f$ as the \emph{objective function}. For an excellent overview of applications of binary programming and equivalent problems we refer to the survey paper by \citeN{boros2002pseudo} and references therein.

The idea to use particle filters for global optimization is not new [\citeNP{del2006sequential}, Section 2.3.1.c], but novel sequential Monte Carlo methodology making use of suitable parametric families on binary spaces \cite{schaefer2011sequential} may allow to construct more efficient samplers for the special case of pseudo-Boolean optimization. We particularly discuss how this methodology connects with the cross-entropy method \cite{Rub:CE1} which is another particle driven optimization algorithm based on parametric families.

The sequential Monte Carlo algorithm as developed by \citeN{schaefer2011sequential} is rather complex compared to local search algorithms such as simulated annealing \cite{kirkpatrick1983optimization} or $k$-opt local search \cite{merz2002greedy} which can be implemented in a few lines. The aim of this paper is to motivate the use of advanced particle methods and sophisticated parametric families in the context of pseudo-Boolean optimization and to provide conclusive numerical evidence that these complicated algorithms can indeed outperform simple heuristics if the objective function has poorly connected strong local maxima. This is not at all clear, since, in terms of computational time, multiple randomized restarts of fast local search heuristics might very well be more efficient than comparatively complex particle approaches.

\subsubsection{Outline}
The article is structured as follows. We first introduce some notation and review how to model an optimization problem \eqref{eq:pb program} as a filtering problem on an auxiliary sequence of probability distributions. Section \ref{sec:smc} describes a sequential Monte Carlo sampler \cite{del2006sequential} designed for global optimization on binary spaces \cite{schaefer2011sequential}. Section \ref{sec:pfbs} reviews three parametric families for sampling multivariate binary data which can be incorporated in the proposed class of particle algorithms. Section \ref{sec:algorithms} discusses how the cross-entropy method \cite{Rub:CE1} and simulated annealing \cite{kirkpatrick1983optimization} can be interpreted as special cases of the sequential Monte Carlo sampler. In Section \ref{sec:applications} we carry out numerical experiments on instances of the unconstrained quadratic binary optimization problem. First, we investigate the performance of the proposed parametric families in particle-driven optimization algorithms. Secondly, we compare variants of the sequential Monte Carlo algorithm, the cross-entropy method, simulated annealing and simple multiple-restart local search to analyze their respective efficiency in the presence or absence of strong local maxima.

\subsubsection{Notation}
We briefly introduce some notation that might be non-standard. We denote scalars in italic type, vectors in italic bold type and matrices in straight bold type. Given a set $M$, we write $\card M$ for the number of its elements and $\ind_M$ for its indicator function. For $a,b\in\Z$ we denote by $\dset{a,b}=\set{a,\dots,b}$ the discrete interval from $a$ to $b$. Given a vector $\v x\in\B^d$ and an index set $I\subseteq\dset{1,d}$, we write $\v x_I\in\B^{\card I}$ for the sub-vector indexed by $I$ and $\v x_{-I}\in\B^{d-\card I}$ for its complement. We occasionally use the norms $\norm{\v x}_{\infty}\eqdef\max_ix_i$ and $\abs{\v x}\eqdef\sum_{i=1}^{d}\abs{x_i}$.

\subsection{Statistical modeling}
\label{sec:stat model}
\subsubsection{Associated probability measures}
For particle optimization, the common approach is defining a family of probability measures $\set{\pi_\varrho\colon\varrho\geq0}$ associated to the optimization problem $\max_{\v x\in\B^{d}} f(\v x)$ in the sense that
\begin{align*}
\pi_0=\uni_{\B^d}, \quad  \lim_{\varrho\to\infty}\pi_\varrho=\uni_{M_f},
\end{align*}
where $\uni_S$ denotes the uniform distribution on the set $S$ and $M_f=\argmax_{\v x\in\B^{d}} f(\v x)$ the set of maximizers. The idea behind this approach is to first sample from a simple distribution, potentially learn about the characteristics of the associated family and smoothly move towards distributions with more mass concentrated in the maxima. We review two well-known techniques to explicitly construct such a family $\pi_\varrho$.

\begin{definition}[Tempered family]
We call $\set{\pi_\varrho\colon\varrho\geq0}$ a tempered family, if it has probability mass functions of the form
\begin{equation}
\label{eq:tempered}
\pi_{\varrho}(\v \gamma)\eqdef\nu_{\varrho}\exp({\varrho}\,f(\v \gamma)),
\end{equation}
where $\nu_{\varrho}^{-1}\eqdef\textstyle\sum_{\v\gamma\in\B^d} \exp({\varrho}\,f(\v \gamma))$.
\end{definition}
As ${\varrho}$ increases, the modes of $\pi_{\varrho}$ become more accentuated until, in the limit, all mass is concentrated on the set of maximizers. The name reflects the physical interpretation of $\pi_{\varrho}(\v x)$ as the probability of a configuration $\v x\in\B^d$ for an inverse temperature ${\varrho}$ and energy function $-f$. This is the sequence used in simulated annealing \cite{kirkpatrick1983optimization}.
\begin{definition}[Level set family]
We call $\set{\pi_\varrho\colon\varrho\geq0}$ a level set family, if it has probability mass functions of the form
\begin{equation}
\label{eq:level set}
\pi_{\varrho}(\v \gamma)\eqdef\card{\ls_{\varrho}}^{-1}\ind_{\ls_{\varrho}}(\v \gamma),
\end{equation}
where $\ls_{\varrho}\eqdef\set{\v \gamma\in\B^d\colon \varrho[f(\v x^{*})-f(\v \gamma)]\leq 1}$ for $\v x^{*}\in M_f$.
\end{definition}
Indeed, $\ls_{\varrho}$ is the super-level set of $f$ with respect to the level $c=f(\v x^{*})-1/{\varrho}$, for $\varrho>0$, and $\pi_{\varrho}(\v \gamma)$ is the uniform distribution on $\ls_{\varrho}$. As ${\varrho}$ increases, the support of $\pi_{\varrho}$ becomes restricted to the points that have an objective value sufficiently close to the maximum of the $f$. In the limit, the support is reduced to the set of global maximizers.

\begin{figure}[ht]
\caption{Associated sequences $\pi_{\varrho_t}$ for a toy example $f\colon\B^3\to[-20,20]$. The colors indicate the advance of the sequences from yellow to red. For simplicity, we choose $\varrho_t=t$ for $t\in\dset{0,16}$.}
\label{fig:sequ}
\subfigure[objective function $f(\v x)$]{
\includegraphics[width=0.3\textwidth]{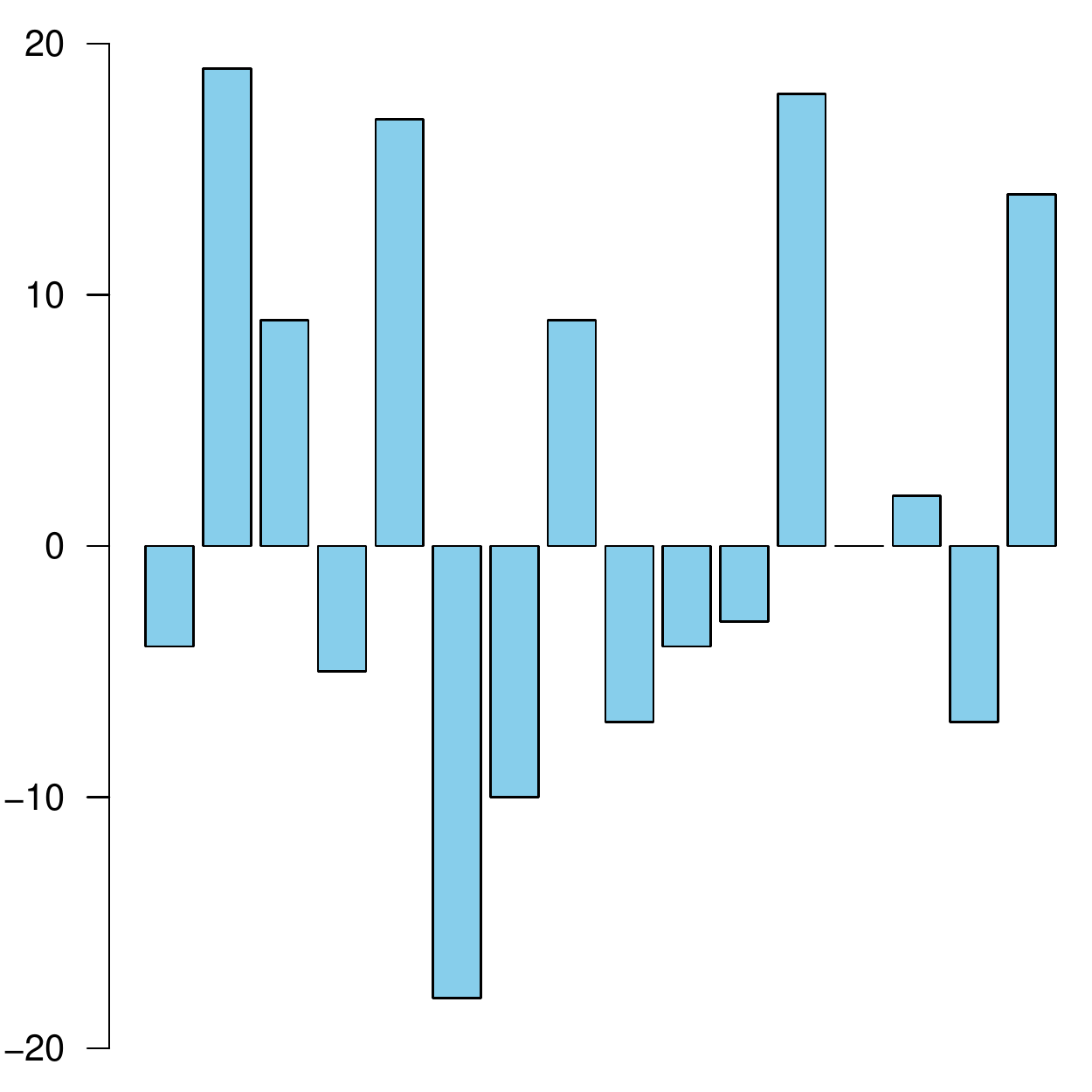}
}
\subfigure[tempered sequence \eqref{eq:tempered}]{
\includegraphics[width=0.3\textwidth]{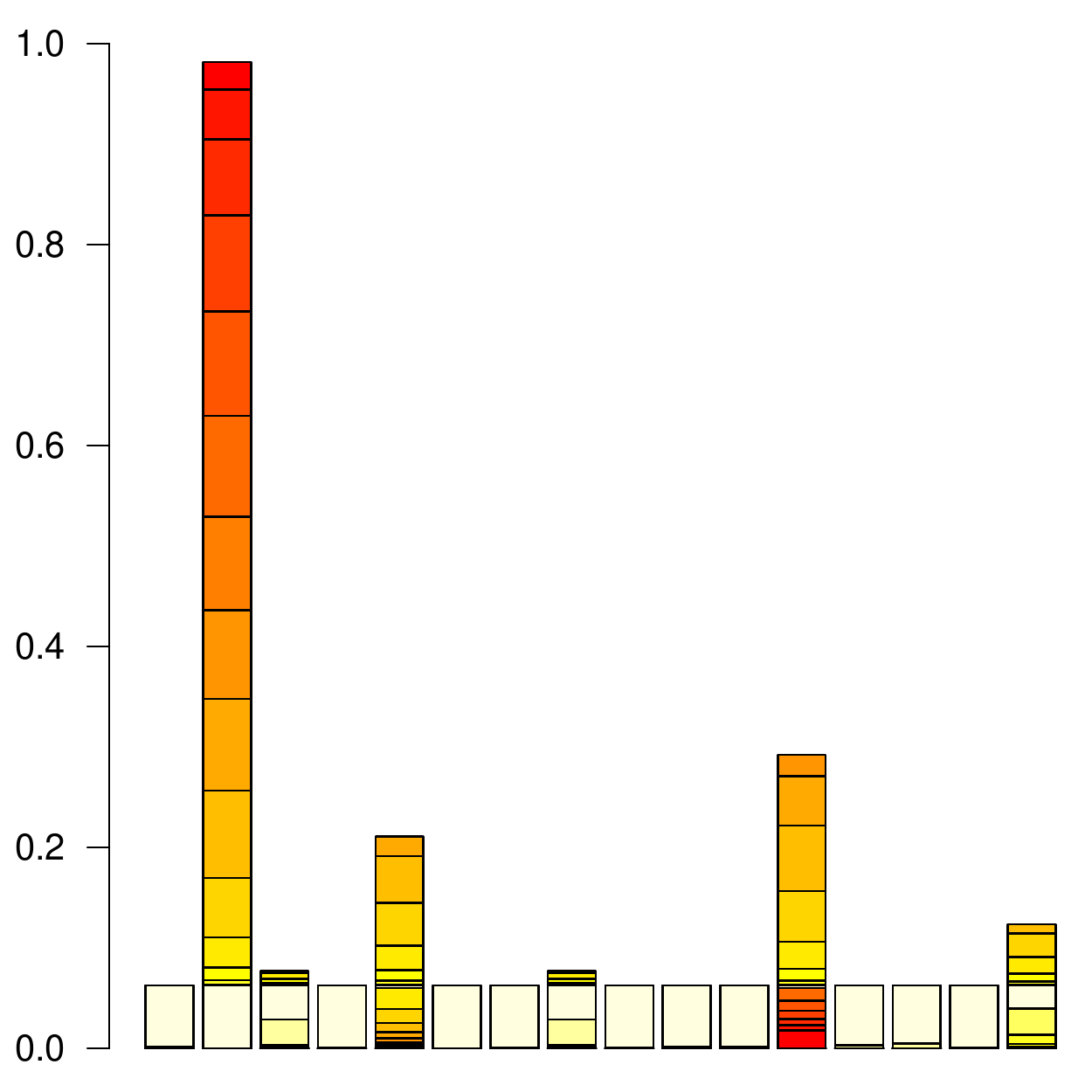}
}
\subfigure[level set sequence \eqref{eq:level set}]{
\includegraphics[width=0.3\textwidth]{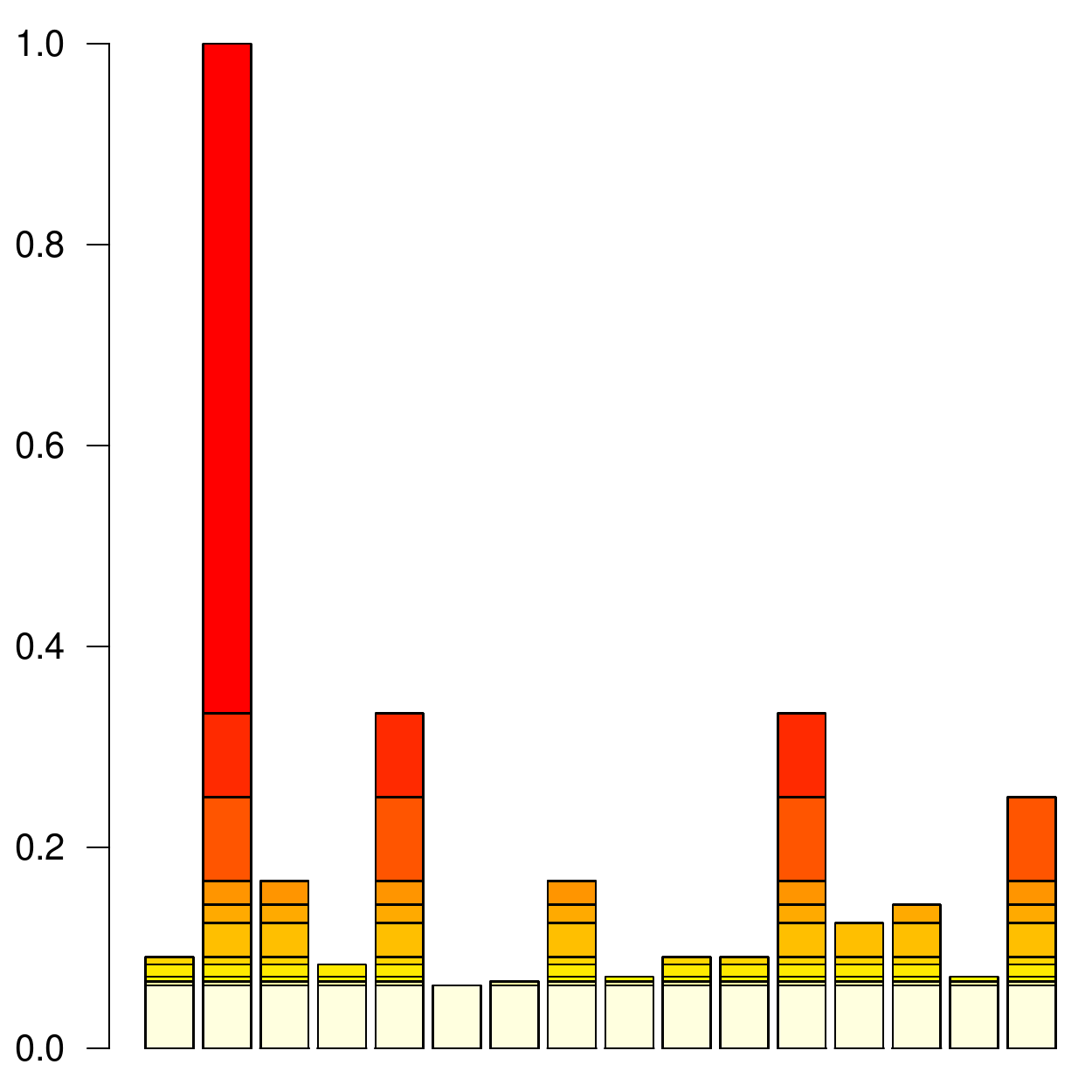}
}
\end{figure}

The particle-driven optimization algorithms are computationally more involved than local search heuristics since we need to construct a sequence of distributions instead of a sequence of states. We shall see that this effort pays off in strongly multi-modal scenarios, where even sophisticated local search heuristics can get trapped in a subset of the state space.

\subsubsection{Rare event simulation}
While the tempered sequence is based on a physical intuition, the level set sequence has an immediate interpretation as a sequence of rare events since, as $\varrho$ increases, the super-level set becomes a `rare event' with respect to the uniform measure. Rare event simulation and global optimization are therefore closely related concepts and methods for rare event estimation can often be adapted to serve as optimization algorithms.

Particle algorithms for rare event simulation include the cross-entropy method \cite{Rub:CE1} and the sequential Monte Carlo sampler \cite{johansen2006sequential}. The former uses the level set sequence, the latter uses a \emph{logistic potential family}
\begin{align*}
\pi_{\varrho}(\v \gamma)\eqdef\nu_{\varrho}\,\logistic(\varrho[f(\v \gamma)-f(\v x^{*})]),
\end{align*}
where $\nu_{\varrho}^{-1}\eqdef\textstyle\sum_{\v\gamma\in\B^d} \logistic(\varrho[f(\v \gamma)-f(\v x^{*})])$ and $\logistic\colon\R\to(0,1),\ \logistic(x)=[1+\exp(-x)]^{-1}$ denotes the logistic function. \citeN{johansen2006sequential} did not specifically design their algorithm for optimization but their approach to static rare event simulation is closely related to the particle optimization framework.

%


%

\section{Sequential Monte Carlo}
\label{sec:smc}
We discuss a static sequential Monte Carlo sampler that uses a transition kernel with independent proposals in the move step based on suitable parametric families. This methodology has been demonstrated to reliably estimate the mean of the posterior distribution in Bayesian variable selection problems for linear normal models \cite{schaefer2011sequential}. In this section, we provide a self-contained description of this framework. For a more general overview of sequential Monte Carlo methods we refer to \citeN{del2006sequential}.

\subsection{Sequential Importance Sampling}
For convenience of notation, we index the sequence of distributions $(\pi_{\varrho_t})_t$ directly by $t$ rather that by the parameter of the family $\varrho_t$. We refer to $\m X=(\v x_1,\dots,\v x_n)\t\in\B^{n\times d}$ and $\v w\in[0,1]^n$ with $\abs{\v w}=1$ as a \emph{particle system} with $n$ particles. We say the particle system $(\v w,\m X)$ \emph{targets} the probability distribution $\pi$ if the empirical distribution $\sum_{k=1}^n w_k\,\delta_{\v x_{k}}$ converges to $\pi$ for $n\to\infty$.

\subsubsection{Importance weights}
We sample $(\v w_0, \m X_0)$ with $\v x_{1,0},\dots,\v x_{n,0} \sim\uni_{\B^d}=\pi_0$ and set $\v w_0=\v 1/n$ to initialize the system. Suppose we are given a particle approximation $(\v w_{t},\m X_{t})$ of $\pi_{t}$ and want to target the subsequent distribution $\pi_{t+1}$. For all $k\in\dset{1,n}$ and $\alpha>0$, we update the weights to
\begin{equation}
\label{eq:imp weights}
w_{k,t,\alpha}\eqdef\frac{u_{k,t,\alpha}}{\sum_{i=1}^n u_{i,t,\alpha}},\text{ where }
u_{k,t,\alpha}\eqdef w_{k,t}\frac{\tilde\pi_{\varrho_{t}+\alpha}(\v x_{k,t})}{\tilde\pi_{\varrho_{t}}(\v x_{k,t})},
\end{equation}
and $\tilde\pi_{\varrho}\propto\pi_{\varrho}$ denotes the unnormalized version of $\pi_{\varrho}$. The normalizing constants $\nu_{\varrho}$ and $\card{\ls_{\varrho}}$ defined in equations \eqref{eq:tempered} and \eqref{eq:level set} are unknown but the algorithm only requires ratios of unnormalized probability mass functions. We refer to $\alpha$ as the \emph{step length} at time $t$. After updating, the particle system targets the distribution
\begin{equation}
\textstyle
\label{eq:emp approx}
\pi_{\varrho_t+\alpha}\approx\sum_{k=1}^n w_{k,t,\alpha}\,\delta_{\v x_{t,k}}.
\end{equation}
As we choose $\alpha$ larger, that is $\pi_{\varrho_t+\alpha}$ further from $\pi_{\varrho_t}$, the weights become more uneven and the accuracy of the importance approximation deteriorates. If we repeat the weighting step, we just increase $\alpha$ and finally obtain an importance sampling estimate of $\pi_\infty=\uni_{M_f}$ with instrumental distribution $\pi_0=\uni_{\B^d}$. This yields a poor estimator since the probability to hit the set $M_f$ with $n$ uniform draws is $1-(1-2^{-d}\card{M_f})^{n}$ and decreases rapidly as the dimension $d$ grows. The pivotal idea behind sequential Monte Carlo is to alternate moderate updates of the importance weights and improvements of the particle system via resampling and Markov transitions.

\begin{definition}[Effective sample size]
The importance weight degeneracy is often measured through the \emph{effective sample size} criterion \cite{KongLiuWong} defined as
\begin{equation*}
\textstyle\eta_n(\v w) \eqdef \left\lbrack n\sum_{k=1}^n w_{k}^2\right\rbrack^{-1}\in[1/n,1].
\end{equation*}
The effective sample size is $1$ if the weights are uniform, that is equal to $1/n$; the effective sample size is $1/n$ if all mass is concentrated in a single particle.
\end{definition}

\subsubsection{Finding the step length}
Given any increasing sequence $(\pi_{\varrho_t})_t$, we could repeatedly reweight and monitor whether the effective sample size falls below a critical threshold. For the special case of annealing via sequential Monte Carlo, however, the effective sample size after weighting $\eta_n(\v w_{t,\alpha})$ is merely a function of $\alpha$. For a particle system $(\m X_{t},\v w_t)$ at time $t$, we pick a step length $\alpha$ such that
\begin{equation}
\label{eq:ess}
\eta_n(\v w_t)\,\beta=\eta_n(\v w_{t,\alpha}),
\end{equation}
that is we lower the effective sample with respect to the current particle approximation by some fixed ratio $\beta\in(0,1)$ [\citeNP{jasra2011inference}, \citeNP{del2011adaptive}]. This ensures a `smooth' transition between two auxiliary distributions, in the sense that consecutive distributions are close enough to approximate each other reasonably well using importance weights; in our numerical experiments, we took $\beta=9/10$. We obtain the associated sequence $(\varrho_t)_t$ by setting $\varrho_{t+1}=\varrho_{t}+\alpha_t$ where $\alpha_t$ is a unique solution of \eqref{eq:ess}. Since $\eta_n(\v w_{t,\alpha})$ is continuous and monotonously decreasing in $\alpha$ we can use bi-sectional search to solve \eqref{eq:ess}.

\subsection{Conditioning the particle system}
In this section, we discuss how to condition the weighted system and improve its quality before we proceed with the next weighting step.

\subsubsection{Resampling}
We replace the system $(\v w_{t+1}, \m X_t)$ targeting $\pi_{t+1}$ by a selection of particles $\hat{\v x}_1,\dots,\hat{\v x}_n$ drawn from the current particle reservoir $\v x_{1,t},\dots,\v x_{n,t}$ such that
\begin{equation*}
\ev{n(\v x_{k})}=n\,w_k,
\end{equation*}
where $n(\v x)$ denotes the number of particles identical with $\v x$. Thus, in the resampled system particles with small weights have vanished while particles with large weights have been multiplied. For the implementation of the resampling step, there exist several recipes. We could apply a multinomial resampling \cite{Gordon} which is straightforward. There are, however, more efficient ways like residual \cite{LiuChen}, stratified \cite{kitagawa1996monte} and systematic resampling \cite{CarClifFearn}. We use the latest in our simulations, see Procedure \ref{algo:resample}.

\begin{algorithm}[ht]
\KwIn{$\v w=(w_1,\dots,w_n),\,\m X=(\v x_1,\dots,\v x_n)\t$}
  $\v v\gets n\,\v w,\ i\gets1,\ c\gets v_1$ \\
  \textbf{sample} $u\sim\uni_{[0,1]}$ \\
\For{$k\in\dset{1,n}$}{
  \lWhile{$c < u$}{$i\gets i+1,\ c \gets c+v_i$}\\
  $\hat{\v x}_k\gets \v x_i,\ u\gets u+1$
}
\Return $\m{\widehat{X}}=(\hat{\v x}_1\dots,\hat{\v x}_n)\t$
\caption{Resampling (systematic)}
\label{algo:resample}
\end{algorithm}

\subsubsection{Moving the system}
\label{sec:move}
If we repeated the weighting and resampling steps several times, we would rapidly reduce the number of different particles to a very few. The key to fighting the depletion of the particle reservoir is moving the particles according to a Markov transition kernel $\kappa_{t+1}$ with invariant measure $\pi_{t+1}$. The particle $\hat{\v x}_{k,t+1}^{(0)}$ is by construction approximately distributed according to $\pi_{t+1}$, and a draw
\begin{equation*}
\hat{\v x}_{k,t+1}^{(1)}\sim\kappa_{t+1}(\bullet\mid\hat{\v x}_{k,t+1}^{(0)})
\end{equation*}
is therefore again approximately distributed according to $\pi_{t+1}$. The last sample of the generated Markov chain
$
(\hat{\v x}_{k,t+1}^{(0)},\dots,\hat{\v x}_{k,t+1}^{(s)})
$
is, for sufficiently many move steps $s\in\N$, almost exactly distributed according to the invariant measure $\pi_{t+1}$ and independent of its starting point.

\subsubsection{Stopping rule}
While we could always apply a fixed number of move steps, we rather use an adaptive stopping criterion based on the number of distinct particles.
\begin{definition}[Particle diversity]
We define the \emph{particle diversity} as
\begin{equation*}
\label{eqn:pd}
\zeta_n(\m X) \eqdef n^{-1}\card{\set{\v x_k\colon k\in\dset{1,n}}}\in[1/n,1].
\end{equation*}
\end{definition}
Ideally, the sample diversity $\zeta_n(\m X)$ should correspond to the expected diversity
\begin{equation*}
\textstyle
\zeta_n(\pi) \eqdef 1\wedge n^{-1}
\sum_{\v\gamma\in\B^d}\ind_{\set{\v x\in\B^d\colon c_n\pi(\v x)\geq 1}}(\v\gamma),
\end{equation*}
where $c_n$ is the smallest value that solves $\sum_{\v\gamma\in\B^d}\lfloor c_n\pi(\v\gamma)\rfloor \geq n$. This is the particle diversity we would expect if we had an independent sample from $\pi(\v x)$. Therefore, if $\kappa_{t+1}$ is fast-mixing, we want to move the system until
\begin{equation*}
\zeta_n(\widehat{\m{X}}\, ^{(s)}_{t+1})\approx\zeta_n(\pi_{t+1}).
\end{equation*}
Since the quantity on the right hand side is unknown, we stop moving the system as soon as the particle diversity reaches a steady state we cannot push it beyond \cite{schaefer2011sequential}.

More precisely, we stop if the absolute diversity is above a certain threshold $\zeta^{*}\approx0.95$ or the last improvement of the diversity is below a certain threshold $\zeta_{\Delta}^{*}>0$. We always stop after a finite number of steps but the thresholds $\zeta^{*}$ and $\zeta_{\Delta}^{*}$ need to be calibrated to the efficiency of the transition kernel. For slow-mixing kernels, we recommend to perform batches of consecutive move steps instead of single move steps.

If the average acceptance rate $\overline\lambda$ of the kernel (see Section \ref{sec:kernels}) is smaller than $\zeta_{\Delta}^{*}$, it is likely that the algorithm stops after the first iteration although further moves would have been necessary. We could adaptively adjust the threshold $\zeta_{\Delta}^{*}$ to be proportional to an estimate of the average acceptance rate; for our numerical experiments, however, we kept it fixed to $\zeta_{\Delta}^{*}\approx 10^{-2}$.

\begin{algorithm}
\DontPrintSemicolon
\KwIn{
\parbox{0.6\textwidth}{
$\m X=(\v x_1^{(0)},\dots,\v x_n^{(0)})\t$ \textbf{ targeting } $\pi$ \\[0.1em]
$\kappa(\v\gamma\mid\bullet)$ \text{ with } $\pi(\v\gamma)=\sum_{\v x\in\B} \pi(\v x) \kappa(\v\gamma\mid\v x)$
}
}
$s\gets 1$ \\
\Repeat{$\zeta(\m X^{(s)})-\zeta(\m X^{(s-1)})<\zeta_{\Delta}^{*}$ \textnormal{\bf or} $\zeta(\m X^{(s)})>\zeta^{*}$}{
\textbf{sample} ${\v x}_k^{(s)}\sim \kappa (\bullet\mid{\v x}_k^{(s-1)})$ \textbf{for all} $k\in\dset{1,n}$ \\
}
\Return $\m X^{(s)}=(\v x_1^{(s)}\dots,\v x_n^{(s)})\t$
\caption{Move}
\label{algo:move}
\end{algorithm}

\subsubsection{Transition kernels}
\label{sec:kernels}
Most transition kernels in Monte Carlo simulations are some variant of the Metropolis-Hastings kernel (see e.g. \citeN{RobCas}),
\begin{equation*}
\kappa_{t+1}\left(\v\gamma\mid\v x\right)
\eqdef\lambda_{q_{t+1}}(\v\gamma,\v x)q_{t+1}(\v\gamma\mid\v x)+
\textstyle
\delta_{\v x}(\v\gamma)\left\lbrack 1-\sum_{\v y\in\B^d}\lambda_{q_{t+1}}(\v y,\v x)q_{t+1}(\v y\mid\v x) \right\rbrack,
\end{equation*}
where we sample from the kernel by proposing a new state $\v\gamma\sim q_{t+1}(\v\gamma\mid \v x)$ and accepting the proposal with probability
\begin{equation}
\label{eq:acc prob}
\lambda_{q_{t+1}}(\v\gamma,\v x)
\eqdef 1\wedge\frac{\tilde\pi_{t+1}(\v\gamma)q_{t+1}(\v x\mid \v\gamma)}{\tilde\pi_{t+1}(\v x)q_{t+1}(\v\gamma\mid \v x)}
\end{equation}
or returning $\v x$ otherwise. Again, we denote by $\tilde\pi_{t+1}\propto\pi_{t+1}$ the unnormalized version of $\pi_{t+1}$ since the kernel only requires the ratio of the unnormalized probability mass functions. 
\begin{definition}[Symmetric kernel]
On binary spaces, a common choice for the proposal distribution is
\begin{equation}
\label{eq:sym kernel}
\textstyle
q(\v\gamma\mid\v x)=\sum_{k=1}^d p_k\delta_k(\abs{\v x-\v\gamma})\,k!(d-k)!/d!,
\end{equation}
with weight vector $\v p\in[0,1]^d$ normalized such that $\abs{\v p}=1$.
\end{definition}
With probability $p_k$, the kernel proposes a uniform draw from the $k$-neighborhood of $\v x$,
\begin{equation}
\label{eq:k neighborhood}
N_k(\v x)\eqdef\set{\v\gamma\in\B^{d}\colon \abs{\v x- \v \gamma}=k}.
\end{equation}
We refer to this type of kernel as \emph{symmetric kernel} since $q(\v\gamma\mid\v x)=q(\v x\mid\v \gamma)$ and equation \eqref{eq:acc prob} simplifies. This class of kernels provide a higher mutation rate than the random-scan Gibbs kernel (see \citeN{schaefer2011sequential} for adiscussion).

Locally operating transition kernels of the symmetric type are known to be slowly mixing. If we put most weight on small values of $k$, the kernel only changes one or a few entries in each step. If we put more weight on larger values of $k$, the proposals will hardly ever be accepted if the invariant distribution $\pi$ is multi-modal. Ideally, we want the particles sampled from the transition kernel to be nearly independent after a few move steps which is often hard to achieve using local transition kernels.

\begin{definition}[Adaptive independent kernel]
For the sequential Monte Carlo algorithm, we use \emph{adaptive independent kernels} which have proposal distributions of the kind
\begin{equation*}
q(\v\gamma\mid\v x)=q_{\theta}(\v\gamma),\quad \theta\in\Theta,
\end{equation*}
which do not depend on the current state $\v x$ but have a parameter $\theta$ which we adapt during the course of the algorithm.
\end{definition}

The adaptive independent kernel is rapidly mixing if we can fit the \emph{parametric family} $q_\theta$ such that the proposal distribution $q_{t+1}=q_{\theta_{t+1}}$ is sufficiently close to the target distribution $\pi_{t+1}$, yielding thus, on average, high acceptance rates $\lambda_{q_{t+1}}$. The general idea behind this approach is to take the information gathered in the current particle approximation into account (see e.g. \citeN{chopin2002sequential}). The usefulness of this strategy for sampling on binary spaces has been illustrated by \citeN{schaefer2011sequential}.

We fit a parameter $\theta_{t+1}$ to the particle approximation of $\pi_{t+1}$ according to some suitable criterion. Precisely, $\theta_{t+1}$ is taken to be the maximum likelihood or method of moments estimator applied to the weighted sample $(\v w_{t+1},\m X_t)$. The choice of the parametric family $q_\theta$ is crucial to the implementation of a sequential Monte Carlo sampler with adaptive independent kernel. We discuss this issue in detail in Section \ref{sec:pfbs}.

Adaptation could, to a certain extent, also be done for local transition kernels. \citeN{nott2005adaptive} propose an adaptive kernel which replaces the full conditional distribution of the Gibbs sampler by an easy to compute linear approximation which is estimated from the sampled particles. This method accelerates Gibbs sampling if the target distribution $\pi$ is hard to evaluate but does not provide fast mixing like the adaptive independent kernel (see \citeN{schaefer2011sequential} for a comparison).

Still, the use of local kernels in the context of the proposed sequential Monte Carlo algorithm might be favorable if, for instance, the structure of the problem allows to rapidly compute the acceptance probabilities of local moves. Further, batches of local moves can be alternated with independent proposals to ensure that the algorithm explores the neighborhood of local modes sufficiently well.

\subsection{Remark on discrete state spaces}
Since the sample space $\B^d$ is discrete, a given particle is not necessarily unique. This raises the question whether it is sensible to store multiple copies of the same weighted particle in our system. In the sequel, we discuss some more details concerning this issue which has only been touched upon briefly by \citeN{schaefer2011sequential}.

Let $n(\v x)$ denote the number of copies of the particle $\v x$ in the system $(\v w,\m X)$. Indeed, for parsimonious reasons, we could just keep a single representative of $\v x$ and aggregate the associated weights to $w_*(\v x)=n(\v x)\,w(\v x)$.

\subsubsection{Impact on the effective sample size}
Shifting weights between identical particles does not affect the nature of the particle approximation but it obviously changes the effective sample size $\eta_n(\v w)$ which is undesirable since we introduced the effective sample size as a criterion to measure the goodness of a particle approximation. From an aggregated particle system, we cannot distinguish the weight disparity induced by reweighting according to the importance function \eqref{eq:imp weights} and the weight disparity induced by multiple sampling of the same states which occurs if the mass of the target distribution is concentrated. More precisely, we cannot tell whether the effective sample size is actually due to the gap between $\pi_t$ and $\pi_{t+1}$ or the presence of particle copies due to the mass of $\pi_t$ concentrating on a small proportion of the state space which occurs by construction of the auxiliary distribution in Section \ref{sec:stat model}.

\subsubsection{Impact on the resample-move step}
Aggregating the weights means that the number of particles is not fixed at runtime. In this case, the straightforward way to implement the move step presented in Section \ref{sec:move} is breaking up the particles into multiple copies corresponding to their weights and moving them separately. But instead of permanently splitting and pooling the weights it seems more efficient to just keep the multiple copies.

We could, however, design a different kind of resample-move algorithm which first augments the number of particles in the move step and then resamples exactly $n$ weighted particles from this extended system using a variant of the resampling procedure proposed by \citeN{fearnhead2003line}. A simple way to augment the number of particles is sampling and reweighting via
\begin{equation*}
\v x_k^{(1)}\sim q_{t+1}(\bullet\mid \v x_k^{(0)}), \quad w_k^{(1)}=w_k\lambda,\ w_k^{(0)}=w_k(1-\lambda),
\end{equation*}
where $\lambda=\lambda_{q_{t+1}}(\v x_{k}^{(1)},\v x_{k}^{(0)})$ denotes the acceptance probability \eqref{eq:acc prob} of the Metropolis-Hastings kernel. We tested this variant but could not see any advantage over the standard sampler presented in the preceding sections. For the augment-resample type algorithm the implementation is more involved and the computational burden significantly higher. In particular, the Rao-Blackwellization effect one might achieve when replacing the accept-reject steps of the transition kernel by a single resampling step does not seem to justify the extra computational effort.

Indeed, aggregating the weights does not only prevent us from using the effective sample size criterion, but also requires extra computational time of $\mathcal O(n\log n)$ in each iteration of the move step since pooling the weights is as complex as sorting. With our application in mind, however, computational time is more critical than memory, and we therefore recommend to refrain from aggregating the weights.

%

%


%

%

\section{Parametric families on binary spaces}
\renewcommand{\algorithmcfname}{Procedure}
\label{sec:pfbs}
We review three parametric families on $\B^d$. In contrast to the similar discussion in \cite{schaefer2011sequential}, we also consider a parametric family which cannot be used in sequential Monte Carlo samplers but in the context of the cross-entropy method. For more details on parametric families on binary spaces we refer to \citeN{schaefer2012logistic}.

\subsection{Suitable parametric families}
\label{sec:properties}
We frame some properties making a parametric family suitable as proposal distribution in sequential Monte Carlo algorithms.
\begin{enumerate}[(a)]
\item For reasons of parsimony, we prefer a family of distributions with at most $d(d+1)/2$ parameters like the multivariate normal.
\item Given a sample $\m X=(\v x_1,\dots,\v x_n)\t$ from the target distribution $\pi$, we need to estimate $\theta^*$ in a reasonable amount of computational time.
\item We need to generate samples $\m Y=(\v y_1,\dots, \v y_m)\t$ from the family $q_\theta$. We need the rows of $\m Y$ to be independent.
\item For the sequential Monte Carlo algorithm, we need to evaluate $q_\theta(\v y)$ point-wise. However, the cross-entropy method still works without this requirement.
\item We want the calibrated family $q_{\theta^*}$ to reproduce e.g. the marginals and covariance structure of $\pi$ to ensure that the parametric family $q_{\theta^*}$ is sufficiently close to $\pi$.
\end{enumerate}

%


%

\subsection{Product family}
The simplest non-trivial distributions on $\B^d$ are certainly those having independent components.

\begin{definition}[Product family]
For a vector $\m m\in(0,1)^d$ of marginal probabilities, we define the \emph{product family}
\begin{equation}
\label{eq:product family}
\textstyle
q^{\Prod}_{\v m}(\v \gamma)\eqdef \prod_{i=1}^d m_i^{\gamma_i}(1-m_i)^{1-{\gamma_i}}.
\end{equation}
\end{definition}

\subsubsection{Properties}
We check the requirement list from Section \ref{sec:properties}:
(a) The product family is parsimonious with $\mathrm{dim}(\theta)=d$.
(b) The maximum likelihood estimator $\hat{\v m}$ is the weighted sample mean.
(c) We can easily sample $\v y\sim q^{\Prod}_{\v m}$.
(d) We can easily evaluate the mass function $q^{\Prod}_{\v m}(\v y)$.
(e) However, the product family does not reproduce any dependencies we might observe in $(\v w,\m X)$.

The last point is the crucial weakness which makes the product family impractical for particle optimization algorithms on strongly multi-modal problems. Consequently, the rest of this section deals with ideas on how to sample binary vectors with a given dependence structure. There are, to our knowledge, two major strategies to this end.
\begin{enumerate}[(1)]
\item We construct a generalized linear model which permits to compute the conditional distributions. We apply the chain rule and write $q_\theta$ as
\begin{equation}
\label{eq:chain rule factorization}
\textstyle
q_{\v\theta}(\v\gamma)=q_{\v\theta}(\v\gamma_1)\prod_{i=2}^d q_{\v\theta}(\v\gamma_i\mid\v\gamma_{1:i-1}),
\end{equation}
which allows to sample the entries of a random vector component-wise.
\item We sample from an auxiliary distribution $\varphi_\theta$ and map the samples into $\B^d$. We call
\begin{equation}
\label{eq:aux distr}
\textstyle
q_{\theta}(\v\gamma)=\int_{\tau^{-1}(\v \gamma)} \varphi_\theta(\v v) d\v v
\end{equation}
a copula family, although we refrain from working with explicit uniform marginals.
\end{enumerate}
We first present a generalized linear model and then review a copula approach.

%


%

\subsection{Logistic conditionals family}
\label{sec:logistic family}
Even for rather simple non-linear models we usually cannot derive closed-form expressions for the marginal probabilities required for sampling according to \eqref{eq:chain rule factorization}. Therefore, we might directly construct a parametric family from its conditional probabilities.

\begin{definition}[Logistic conditionals family]
We define, for a lower triangular matrix $\m A\in\R^{d\times d}$, the \emph{logistic conditionals family} as
\begin{align*}
\label{eq:lb}
q^{\LogCo}_{\m A}(\v \gamma)
&
\eqdef
\textstyle
\prod_{i\in \dset{1,d}}\logistic\left(a_{ii}+\sum_{j=1}^{i-1} a_{ij}\gamma_j\right)^{\gamma_i}
\left[1-\logistic\left(a_{ii}+\sum_{j=1}^{i-1} a_{ij}\gamma_j\right)\right]^{1-\gamma_i}
\end{align*}
where $\logistic\colon\R\to(0,1),\ \logistic(x)=[1+\exp(-x)]^{-1}$ is the logistic function. We readily identify the product family $q^{\Prod}_{\v m}$ as the special case $\m A=\diag{\logistic^{-1}(\v m)}$.
\end{definition}

The virtue of the logistic conditionals family is that, by construction, we can sample a random vector component-wise while the full probability $q_{\m A}^{\LogCo}(\v y)$ of the sample $\v y$ is computed as a by-product of Procedure \ref{algo:sampling}. We refer to the Online Supplement for instructions on how to fit the parameter $\m A$.

\begin{algorithm}
$\v y=(0,\dots,0),\ p\gets 1$ \\
\For{$i\in\dset{1,d}$}{
   $r\gets q_{\m A}^{\LogCo}(y_i=1\mid\v y_{1:i-1})=\logistic({a_{ii}+\sum_{j=1}^{i-1}a_{ij}y_j})$ \\
   $u\sim\mathcal{U}[0,1]$ \\
   \bf{ if } $u<r$ \bf{ then }$y_i\gets1$ \\
   $p\gets\begin{cases}
	    p\cdot r    & \textbf{if }\ \ y_i=1 \\
	    p\cdot (1-r) & \textbf{if }\ \ y_i=0
            \end{cases}$ \\
}
\Return $\v y,\ p$
\caption{Sampling via chain rule factorization}
\label{algo:sampling}
\end{algorithm}

\subsubsection{Properties}
We check the requirement list from Section \ref{sec:properties}:
(a) The logistic conditionals family is sufficiently parsimonious with $\mathrm{dim}(\theta)=d(d+1)/2$.
(b) We can fit the parameter $\m A$ via likelihood maximization. The fitting is computationally intensive but feasible.
(c) We can sample $\v y\sim q^{\LogCo}_{\m A}$ using the chain rule factorization \eqref{eq:chain rule factorization}.
(d) We can exactly evaluate $q^{\LogCo}_{\m A}(\v y)$.
(e) The family $q^{\LogCo}_{\m A}$ reproduces the dependency structure of the data $\m X$ although we cannot explicitly compute the marginal probabilities.

\subsection{Gaussian copula family}
\label{sec:Gaussian copula}
Let $\varphi_\theta$ be a family of multivariate auxiliary distributions on $\mathbb X$ and $\tau\colon\mathbb X \to \B^d$ a mapping into the binary space. We can sample from the copula family \eqref{eq:aux distr} by setting $\v x=\tau(\v v)$ for a draw $\v v\sim \varphi_\theta$ from the auxiliary distribution. Most multivariate parametric families with at most $d(d+1)/2$ parameters appear to either have a rather limited dependency range or they do not scale to higher dimensions \cite{joe1996families}. Therefore, the natural and seemingly only viable option for $\varphi_\theta$ is the multivariate normal distribution \cite{emrich1991method}.

\begin{definition}[Gaussian copula family]
For a vector $\v a\in\R^{d}$ and a correlation matrix $\m \Sigma\in\R^{d\times d}$, we introduce the mapping
\begin{equation*}
\tau_{\v a}\colon\R^{d}\to\B^{d},\
\tau_{\v a}(\v v)\eqdef(\ind_{(-\infty,a_i]}(v_1),\dots,\ind_{(-\infty,a_d]}(v_d)),
\end{equation*}
and define the \emph{Gaussian copula family} as
\begin{equation*}
\textstyle
q^{\Gau}_{\v a,\m \Sigma}(\v\gamma)
\eqdef(2\pi)^{-\frac{d}{2}}\det{\m \Sigma}^{-\frac{1}{2}}\textstyle\int_{\tau_{\v a}^{-1}(\v\gamma)}\exp\left(-\frac{1}{2}\,\v v\t\m \Sigma^{-1}\v v\right)\,d\v v.
\end{equation*}
\end{definition}
For index sets $I\subseteq \dset{1,d}$, the cross-moments
\begin{equation*}
m_I=\textstyle\sum_{\v\gamma\in\B^{d}}q^{\Gau}_{\v a,\m \Sigma}(\v\gamma)\prod_{i\in I}\gamma_i
\end{equation*}
are equal the cumulative distribution function of the multivariate normal with respect to the entries indexed by $I$ (see \citeN{schaefer2012logistic} for a more detailed discussion). In particular, the first and second moments are
\begin{equation*}
m_i=\varPhi_1(a_i),\quad m_{ij}=\varPhi_2(a_i,a_j;\sigma_{ij}),\quad i,j\in\dset{1,d},
\end{equation*}
where $\varPhi_1(\cdot)$ and $\varPhi_2(\cdot,\cdot;\sigma_{ij})$ denote the cumulative distribution functions of the univariate and bivariate normal distributions with zero mean, unit variance and correlation coefficient $\sigma_{ij}\in[-1,1]$. We refer to the Online Supplement for instructions on how to fit the parameters $\v a$ and $\m\Sigma$.

\subsubsection{Properties}
We check the requirement list from Section \ref{sec:properties}:
(a) The Gaussian copula family is sufficiently parsimonious with $\mathrm{dim}(\theta)=d(d+1)/2$.
(b) We can fit the parameters $\v a$ and $\m \Sigma$ via method of moments. However, the parameter $\m \Sigma$ is not always positive definite.
(c) We can sample $\v y\sim q^{\Gau}_{\v a,\m \Sigma}$ using $\v y=\tau_{\v a}(\v v)$ with $\v v\sim \varphi_{\m \Sigma}$.
(d) We cannot easily evaluate $q^{\Gau}_{\v a,\m \Sigma}(\v y)$ since this requires computing high-dimensional integral expressions which is a computationally challenging problem in itself (see e.g. \citeN{genz2009computation}). The Gaussian copula family is therefore less useful for sequential Monte Carlo samplers but can be incorporated into the cross-entropy method reviewed in Section \ref{sec:cross entropy}.
(e) The family $q^{\Gau}_{\v a,\m \Sigma}$ reproduces the exact mean and, possibly scaled, correlation structure.

\subsection{Toy example}
We briefly discuss a toy example to illustrate the usefulness of the parametric families. For the quadratic function
\begin{equation}
\label{eq:toy exa}
f(\v x)=\v x\t\m F\v x,\quad \m F\eqdef
   \begin{pmatrix}
      1&  2&  1&  0 \\
      2&  1& -3& -2 \\
      1& -3&  1&  2 \\
      0& -2&  2& -2      
   \end{pmatrix},
\end{equation}
the associated probability mass function $\pi(\v\gamma)\propto\exp(\v\gamma\t\m F\v\gamma)$ has a correlation matrix
\begin{equation*}
\m R\approx\begin{pmatrix}
       1&       0.127&  -0.106&  -0.101 \\
       0.127&       1&  -0.941&  -0.866 \\
      -0.106&  -0.941&       1&   0.84  \\
      -0.101&  -0.866&    0.84&   1    
     \end{pmatrix},
\end{equation*}
which indicates that this distribution has considerable dependencies and its mass function is therefore strongly multi-modal. We generate pseudo-random data from $\pi$, adjust the parametric families to the data and plot the mass functions of the fitted parametric families.

\begin{figure}[ht]
\caption{Toy example showing how well the parametric families replicate the mass function of the distribution $\pi(\v\gamma)\propto\exp(\v\gamma\t\m F\v\gamma)$ as defined in \eqref{eq:toy exa}.}
\label{fig:toy exa}
\begin{center}
\subfigure[True mass function $\pi(\v\gamma)$]{
\includegraphics[width=0.46\textwidth]{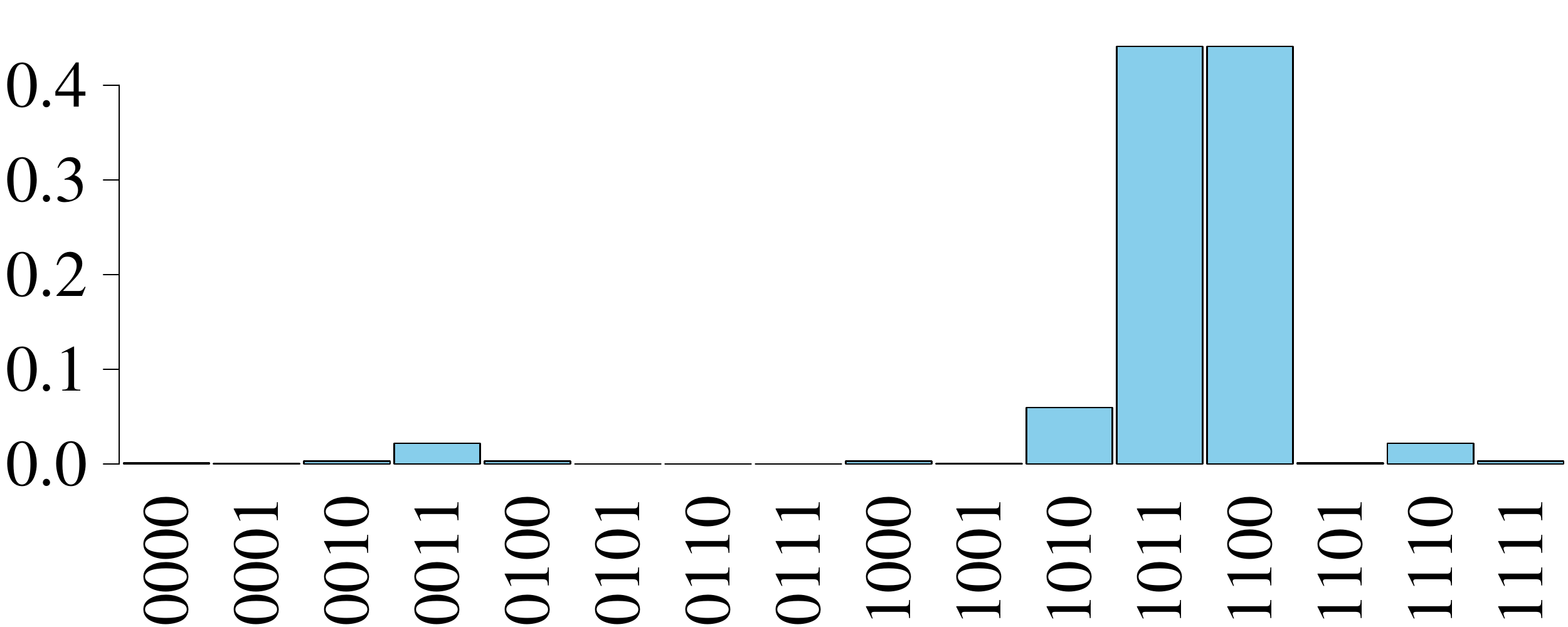}
}
\subfigure[Product family $q_{\v m}(\v\gamma)$]{
\includegraphics[width=0.46\textwidth]{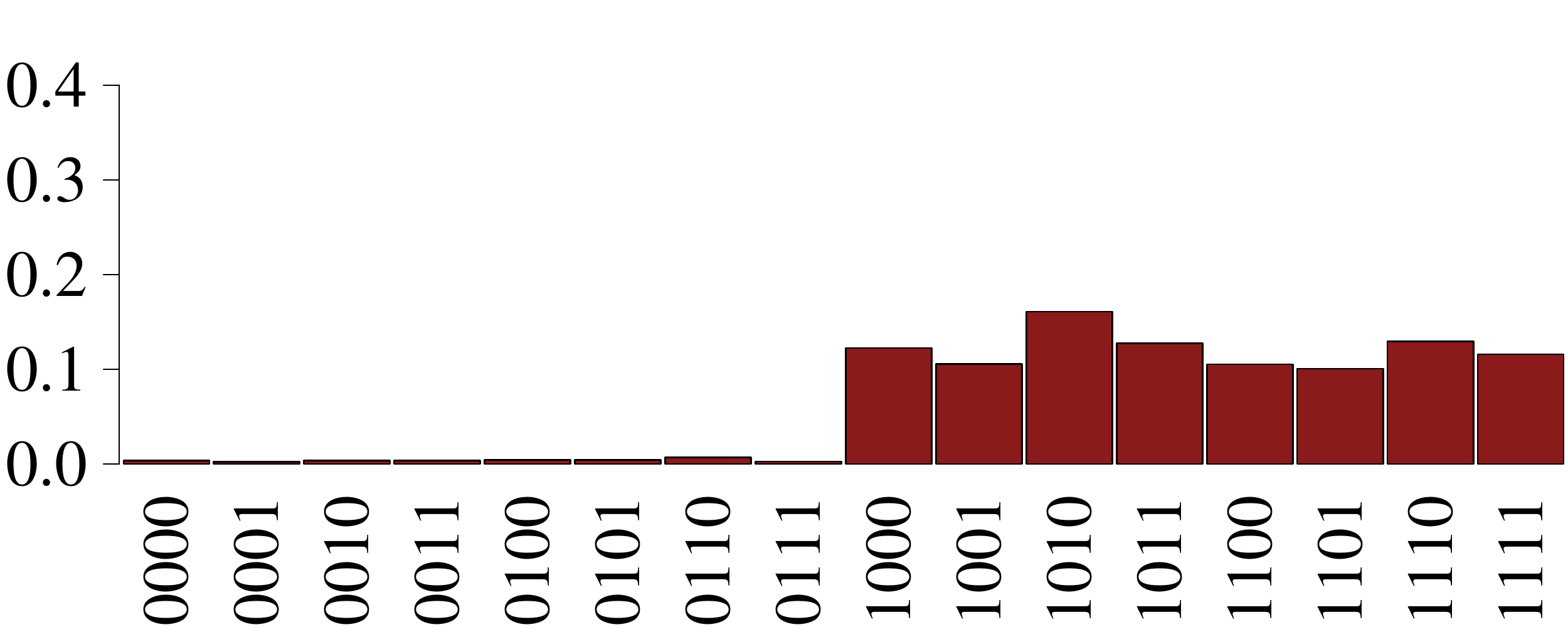}
}
\subfigure[Logistic conditionals family $q_{\m A}(\v\gamma)$]{
\includegraphics[width=0.46\textwidth]{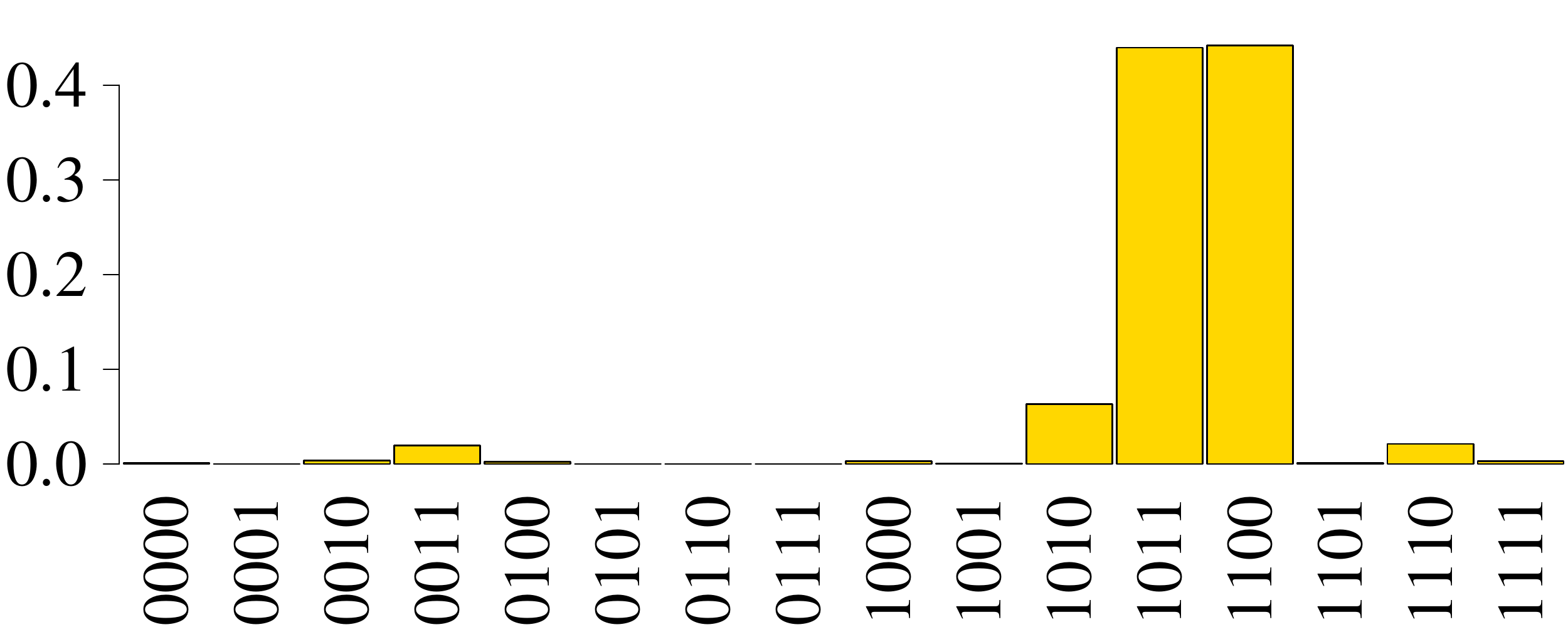}
}
\subfigure[Gaussian copula family $q_{\v a, \m\Sigma}(\v\gamma)$]{
\includegraphics[width=0.46\textwidth]{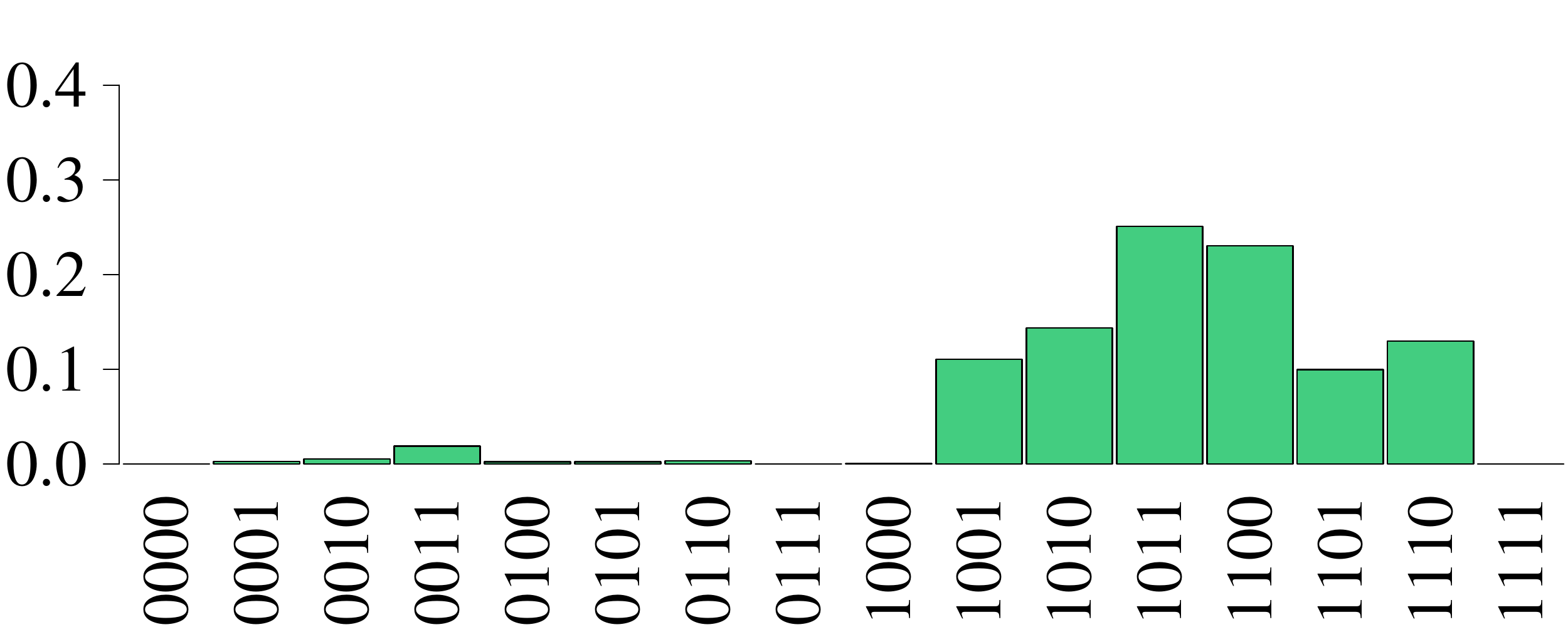}
}
\end{center}
\end{figure}

Figure \ref{fig:toy exa} shows how the three parametric families cope with reproducing the true mass function. Clearly, the product family is not close enough to the true mass function to yield a suitable instrumental distribution while the logistic conditional family almost copies the characteristics of $\pi$ and the Gaussian copula family allows for an intermediate goodness of fit.

%

%


%

%

\section{Optimization algorithms}
\label{sec:algorithms}
In this section, we provide a synopsis of all steps involved in the sequential Monte Carlo algorithm and connect this framework to the cross-entropy method and simulated annealing. In Table \ref{tab:seq}, we state the necessary formulas for the tempered and the level set sequence introduced in Section \ref{sec:stat model}.

\subsection{Sequential Monte Carlo}
For convenience, we summarize the complete sequential Monte Carlo sampler in Algorithm \ref{algo:smc}. Note that, in practice, the sequence $\pi_{\varrho_t}$ is not indexed by $t$ but rather by $\varrho_t$, which means that the counter $t$ is only given implicitly.

The algorithm terminates if the particle diversity sharply drops below some threshold $\delta>0$ which indicates that the mass has concentrated in a single mode. If we use a kernel with proposals from a parametric family $q_{\theta_t}$, we might already stop if the family degenerates in the sense that only a few components of $q_{\theta_t}$, say less than $d^{*}=12$, are random while the others are constant ones or zeros. In this situation, additional moves using a parametric family are a pointless effort. We either return the maximizer within the particle system or we solve the subproblem of dimension $d^*$ by brute force enumeration. We might also perform some final local moves in order to further explore the regions of the state space the particles concentrated on.

\renewcommand{\algorithmcfname}{Algorithm}
\setcounter{algocf}{0}
\begin{algorithm}
\KwIn{$f\colon \B^d\to\R$}
\textbf{sample} $\v x_k\stackrel{\mathrm{iid}}{\sim}\uni_{\B^d}$ \textbf{for all} $k\in\dset{1,n}$. \\[0.3em]
$\alpha\gets\textbf{find step length}(0,\m X)$, $\v w\gets\textbf{importance weights}(\alpha,\pi,\m X)$ \\[0.3em]
\While{$\zeta_{n}(\m X)>\delta$}{$ $\\[0.3em] 
\begin{tabular}{lll}
$q_\theta	$&\hspace{-3mm}$\gets\textbf{fit parametric family}(\v w, \m X)$ 	&\hspace{-3mm} (see Section \ref{sec:pfbs}) \\[0.3em]
$\m{\widehat{X}}	$&\hspace{-3mm}$\gets\textbf{resample}(\v w,\m X)$ 		&\hspace{-3mm} (Procedure \ref{algo:resample}) \\[0.3em]
$\m X		$&\hspace{-3mm}$\gets\textbf{move}(\kappa_{\pi, q_\theta}, \m{\widehat{X}})$&\hspace{-3mm} (Procedure \ref{algo:move}) \\[0.3em]
$\alpha		$&\hspace{-3mm}$\gets\textbf{find step length}(\varrho, \m X)$		& \\[0.3em]
$\v w		$&\hspace{-3mm}$\gets\textbf{importance weights}(\alpha,\pi_{\varrho},\m X)$	& \\[0.3em]
$\varrho	$&\hspace{-3mm}$\gets\varrho+\alpha$&
\end{tabular}
}
\Return $\argmax_{\v x\in\set{\v x_1,\dots,\v x_n}} f(\v x)$
\caption{Sequential Monte Carlo optimization}
\label{algo:smc}
\end{algorithm}

\subsection{Cross-entropy method}
\label{sec:cross entropy}
For the level set sequence, the effective sample size is the fraction of the particles which have an objective function value greater than $\max_{\v x\in\B^d}f(\v x)-1/(\varrho_t+\alpha)$; see Table \ref{tab:seq} and equation \eqref{eq:level set}. The remaining particles are discarded since their weights equal zero. Consequently, there is no need to explicitly compute $\alpha_t$ as a solution of \eqref{eq:ess}. We simply order the particles $\v x_k$ according to their objective values $f(\v x_k)$ and only keep the $n(1-\beta)$ particles with the highest objective values.

Rubinstein \citet{Rub:CE1}, who popularizes the use of level set sequences in the context of the cross-entropy method, refers to $n(1-\beta)$ as the size of the \emph{elite sample}. The cross-entropy method has been applied successfully to a variety of combinatorial optimization problems, some of which are equivalent to pseudo-Boolean optimization \cite{Rub:bookCE}, and is closely related to the proposed sequential Monte Carlo framework.

However, the central difference between the cross-entropy method and the sequential Monte Carlo algorithm outlined above is the use of the invariant transition kernel in the latter. We obtain the cross-entropy method as a special case if we replace the kernel $\kappa_t$ by its proposal distribution $q_{\theta_t}$. The sequential Monte Carlo approach uses a smooth family of distributions $\set{\pi_\varrho\colon\varrho\geq0}$ and explicitly schedules the evolution $\pi_{\varrho_t}$ which in turn leads to the proposal distributions $q_{\theta_t}$. The cross-entropy method, in contrast, defines the subsequent proposal distribution
\begin{equation*}
q_{\theta_{t+1}}\approx q_{\theta_{t}}\ind_{\ls_{\varrho_{t+1}}} 
\end{equation*}
without any reference sequence $\pi_t$ to balance the speed of the particle evolution.

In order to decelerate the advancement of the cross-entropy method, we introduce a lag parameter $\tau\in[0,1)$ and use a convex combination of the previous parameter $\theta_{t-1}$ and the parameter $\hat\theta_t$ fit to the current particle system, setting
\begin{equation*}
\theta_{t}\eqdef(1-\tau)\hat\theta_{t}+\tau\theta_{t-1}.
\end{equation*}
However, there are no guidelines on how to adjust the lag parameter during the run of the algorithm. Therefore, the sequential Monte Carlo algorithm is easier to calibrate since the reference sequence $\pi_t$ controls the stride and automatically prevents the system from overshooting.

On the upside, the cross-entropy method allows for a broader class of auxiliary distributions $q_{\theta_t}$ since we do not need to evaluate $q_{\theta_t}(\v x)$ point-wise which is necessary in the computation of the acceptance probability of the Hastings kernel; see Section \ref{sec:Gaussian copula}.

\begin{table}
\caption{Formulas for optimization sequences}
\label{tab:seq}
\begin{center}
\begin{tabular}{l|c|c}
& $\exp(\varrho f)$
& $\ind_{\ls_{\varrho}}$ \\[1mm]
\hline
&&\\[-3mm]
$u_{t,\alpha}(\v x_{k,t})$
& $\displaystyle e^{\alpha f(\v x_{k,t})}$
& $\displaystyle \ind_{\ls_{\varrho_{t}+\alpha}}(\v x_{k,t})$ \\[5mm]
$\eta_n(\v w_{t,\alpha})$ 
& $\displaystyle\frac{\left\lbrack\sum_{k=1}^n e^{\alpha f(\v x_{k,t})}\right\rbrack^2}{n\sum_{k=1}^n e^{2\alpha f(\v x_{k,t})}}$
& $\displaystyle\frac{\card{\set{\v x_{k,t}\mid k\in\dset{1,n}} \cap \ls_{\varrho_{t}+\alpha}}}{n}$ \\[5mm]
$\lambda_{q_{t+1}}(\v\gamma\mid\v x_{k,t})$
& $\displaystyle 1\wedge \frac{e^{\alpha(f(\v\gamma)-f(\v x_{k,t}))}}{e^{\log q_t(\v \gamma)-\log q_t(\v x_{k,t})}}$
& $\displaystyle 1\wedge\frac{\ind_{\ls_{\varrho_{t+1}}}(\v \gamma)}{e^{\log q_t(\v \gamma)-\log q_t(\v x_{k,t})}}$
\end{tabular}
\end{center}
\end{table}

\subsection{Simulated annealing}
\label{sec:sim ann}
A well-studied approach to pseudo-Boolean optimization is simulated annealing \cite{kirkpatrick1983optimization}.
While the name stems from the analogy to the annealing process in metallurgy, there is a pure statistical meaning to this setup. We can picture simulated annealing as approximating the mode of a tempered sequence \eqref{eq:tempered} using a single particle. Since a single observation does not allow for fitting a parametric family, we have to rely on symmetric transition kernels \eqref{eq:sym kernel} in the move step.

A crucial choice is the sequence $\varrho_t$ which in this context is often referred to as the \emph{cooling schedule}. There is a vast literature advising on how to calibrate $\varrho_t$ where a typical guideline is the expected acceptance rate of the Hastings kernel. We calibrate $\varrho_t$ such that the empirical acceptance rate
\begin{equation*}
\textstyle
\overline{\lambda}_{t-s:t}\eqdef\sum_{r=t-s}^{t}\lambda_r,\quad s>0
\end{equation*}
follows approximately $(t+1)^{-5}$ for $t\in[0,1]$. There are variants of simulated annealing which use more complex cooling schedules, tabu lists and multiple restarts, but we stick to this simple version for the sake of simplicity. Algorithm \ref{algo:sim ann} describes the version we use in our numerical experiments in Section \ref{sec:perf algo}.


\begin{algorithm}[ht]
\KwIn{$f\colon \B^d\to\R,\,T^{*}\in\R$}
$\v x\sim\uni_{\B^d},\, \v x^*\gets\v x,\,t\gets0,\, T_{\Delta}\gets0$ (time elapsed) \\
\While{$T_{\Delta}<T^{*}$}{
  \textbf{sample }$\v\gamma\sim \uni_{N_1(\v x)},\ u\sim\uni_{[0,1]},\ \lambda_t\gets1\wedge \exp\left[\varrho_t\,(f(\v \gamma)-f(\v x))\right]$ \\
  \lIf{$u<\lambda_t$}{$\v x \gets \v\gamma$} \\
  \lIf{$f(\v x) > f(\v x^*)$}{$\v x^* \gets \v x$} \\
  \textbf{adjust }$\varrho_t$ \textbf{ such that } $\overline{\lambda}_{t-s:t}\approx(1+T_{\Delta}/T^{*})^{-5}$ \\
  $t\gets t+1$
}
\Return $\v x^*$
\caption{Simulated annealing optimization}
\label{algo:sim ann}
\end{algorithm}

\subsection{Randomized local search}
\label{sec:local search}
We describe a greedy local search algorithm which works on any state space that allows for defining a neighborhood structure. The typical neighborhood on binary spaces is the $k$-neighborhood defined in \eqref{eq:k neighborhood}. A greedy local search algorithm computes the objective value of all states in the current neighborhood and moves to the best state found until a local optimum is reached. The local search algorithm is called $k$-opt if it searches the neighborhood $\cup_{i=1}^{k}N_i(\cdot)$ (see e.g. \citeN{merz2002greedy} for a discussion).

The algorithm can be randomized by repeatedly restarting the procedure from randomly drawn starting points. There are more sophisticated versions of local search algorithms exploit the properties of the objective function but even a simple local search procedure can produce good results \cite{alidaee2010theorems}. Algorithm \ref{algo:local search} describes the $1$-opt local search procedure we use in our numerical experiments in Section \ref{sec:perf algo}.

\begin{algorithm}[ht]
\KwIn{$f\colon \B^d\to\R,\,T^{*}\in\R$}
$\v x^{*}\sim\uni_{\B^d},\, T_{\Delta}\gets0$ (time elapsed) \\
\While{$T_{\Delta}<T^{*}$}{
  $\v x\sim\uni_{\B^d}$ \\
  \While{$\v x$\textnormal{\textbf{ is not a local optimum}}}{
    $\v x\gets\argmax_{\v \gamma\in N_1(\v x)} f(\v \gamma)$\\
  }
  \lIf{$f(\v x) > f(\v x^*)$}{$\v x^* \gets \v x$}\\
}
\Return $\v x^*$
\caption{Randomized local search}
\label{algo:local search}
\end{algorithm}

%


%

\section{Applications}
\label{sec:applications}


\subsection{Unconstrained Quadratic Binary Optimization}
\subsubsection{Introduction}
It is well-known that any pseudo-Boolean function $f\colon\B^{d}\to\R$ can be written as a multi-linear function
\begin{align}
\label{eq:multi-linear}
f(\v x)
=\sum_{I\subseteq \dset{1,d}} f\left(\ind_I(1), \dots,\ind_I(d)\right)\prod_{i\in I}x_i\prod_{i\in \dset{1,d}\setminus I}(1-x_i)
=\sum_{I\subseteq \dset{1,d}}  a_I \prod_{i\in I}x_i,
\end{align}
where $a_I\in\R$ are real-valued coefficients. We say the function $f$ is of order $k$ if the coefficients $a_I$ are zero for all $I\subseteq \dset{1,d}$ with $\card I > k$. While optimizing a first order function is trivial, optimizing a non-convex second order function is already an NP-hard problem \cite{garey1979guide}. 

In the sequel, we focus on optimization of second order pseudo-Boolean functions to exemplify the stochastic optimization schemes discussed in the preceding sections. If $f$ is a second order function, we restate program \eqref{eq:pb program} as
\begin{equation}
\label{eq:ubqo program}
\begin{tabular}{ll}
\text{maximize }  & $\v x\t \m F \v x$ \\[0.2em]
\text{subject to} & $ \v x\in\B^d$,
\end{tabular}
\end{equation}
where $\m F\in\R^{d\times d}$ is a symmetric matrix. We call \eqref{eq:ubqo program} an unconstrained quadratic binary optimization problem (\textsc{uqbo}); we refer to \citeN{boros2007local} for a list of applications and equivalent problems. In the literature it is also referred to as unconstrained quadratic Boolean or bivalent or zero-one programming \cite{beasley1998heuristic}.

\subsubsection{Particle optimization and meta-heuristics}
\label{sec:meta}
Meta-heuristics are a class of algorithms that optimize a problem by improving a set of candidate solutions without systematically enumerating the state space; typically they deliver solutions in polynomial time while an exact solution has exponential worst case running time. The outcome is neither guaranteed to be optimal nor deterministic since most meta-heuristics are randomized algorithms. We briefly discuss the connection to particle optimization against the backdrop of the unconstrained quadratic binary optimization problem where we roughly separate them into two classes: local search algorithms and particle-driven meta-heuristics.

Local search algorithms iteratively improve the current candidate solution through local search heuristics and judicious exploration of the current neighborhood; examples are local search \cite{boros2007local,merz2002greedy}, tabu search \cite{glover1998adaptive,palubeckis2004multi}, simulated annealing \cite{katayama2001performance}. Particle driven meta-heuristics propagate a set of candidate solutions and improve it through recombination and local moves of the particles; examples are genetic algorithms \cite{merz1999genetic}, memetic algorithms \cite{merz2004memetic}, scatter search \cite{amini1999scatter}. For comparisons of these methods we refer to \citeN{hasan2000comparison} or \citeN{beasley1998heuristic}.

The sequential Monte Carlo algorithm and the cross-entropy method are clearly in the latter class of particle-driven meta-heuristics. The idea behind sequential Monte Carlo is closely related to the intuition behind population (or swarm) optimization and genetic (or evolutionary) algorithms. However, the mathematical framework used in sequential Monte Carlo allows for a general formulation of the statistical properties of the particle evolution while genetic algorithms are often problem-specific and empirically motivated.

\subsubsection{Particle optimization and exact solvers}
If we can explicitly derive the multi-linear representation \eqref{eq:multi-linear} of the objective function, there are techniques to turn program \eqref{eq:pb program} into a linear program. For the \textsc{uqbo} it reads
\begin{equation}
\label{eq:lin ubqo}
\begin{tabular}{ll}
\text{maximize }  & $\displaystyle f(\v x)=2\sum_{i=1}^d\sum_{j=1}^{i-1} f_{ij}x_{ij}+\sum_{i=1}^df_{ii}x_{ii}$ \\[1.5em]
\text{subject to} & $\v x\in\B^{d(d+1)/2}$ \\
		  & $\left.
			\begin{array}{l}
			      \hspace{-1ex}x_{ij}\leq x_{ii} \\
			      \hspace{-1ex}x_{ij}\leq x_{jj} \\
			      \hspace{-1ex}x_{ij}\geq x_{ii} + x_{jj} -1 \\
			\end{array}
		    \right\}\text{ for all}\ i,j\in \dset{1,d}$.
\end{tabular}
\end{equation}
Note, however, that there are more parsimonious linearization strategies than this straightforward approach [\citeNP{hansen2009improved}, \citeNP{gueye2009linear}]. The transformed problem allows to access the tool box of linear integer programming which consist of branch-and-bound algorithms that are combined with rounding heuristics, various relaxations techniques and cutting plane methods [\citeNP{pardalos1990computational}, \citeNP{palubeckis1995heuristic}].

Naturally, the question arises whether particle-driven meta-heuristics can be incorporated into exact solvers to improve branch-and-bound algorithms. Indeed, stochastic meta-heuristics deliver lower bounds for maximization problems, but particle-driven algorithms are computationally somewhat expensive for this purpose unless the objective function is strongly multi-modal and other heuristics fail to provide good results; see the discussion in Section \ref{sec:extreme}.

However, the sequential Monte Carlo approach in combination with the level set sequence \eqref{eq:level set} might also be useful to determine a global branching strategy, since the algorithm provides an estimator for
\begin{equation*}
\textstyle
\overline{\v\gamma}_{c}\eqdef\card{\ls_{c}}^{-1}\sum_{\v\gamma\in\B^d}\v\gamma\,\ind_{\ls_{c}}(\v\gamma),
\end{equation*}
which is the average of the super-level set $\ls_{c}\eqdef\set{\v x\in\B^d\colon f(\v x)\geq c}$. These estimates given for a sequence of levels $c$ might provide branching strategies than are superior to local heuristics or branching rules based on fractional solutions. A further discussion of this topic is beyond the scope of this paper but it certainly merits consideration.


\subsection{Construction of test problems}
\label{sec:test problems}
\subsubsection{Introduction}
The meta-heuristics we want to compare do not exploit the quadratic structure of the objective function and might therefore be applied to any binary optimization program. If the objective function can be written in multi-linear form like \eqref{eq:ubqo program} there are efficient local search algorithms \cite{boros2007local,merz2002greedy} which exploit special properties of the target function and easily beat particle methods in terms of computational time.

Therefore, the use of particle methods is particularly interesting if the objective function is expensive to compute or even a black box. The posterior distribution in Bayesian variable selection for linear normal models is an example of such an objective function (see \citeN{schaefer2011sequential} and references therein). We stick to the \textsc{uqbo} for our numerical comparison since problem instances of varying difficulty are easy to generate and interpret while the results carry over to general binary optimization.

In the vast literature on \textsc{uqbo}, authors typically compare the performance of meta-heuristics on a suite of randomly generated problems with certain properties. \citeN{pardalos1991construction} proposes standardized performance tests on symmetric matrices $\m F\in\Z^{d\times d}$ with entries $f_{ij}$ drawn from the uniform
\begin{equation*}
q_{c}(k)\eqdef\frac{1}{2c}\ind_{\dset{-c,c}}(k), \quad c\in\N.
\end{equation*}
The test suites generated by \citet[\href{http://people.brunel.ac.uk/~mastjjb/jeb/orlib/bqpinfo.html}{OR-library}]{beasley1990or} and \citeN{glover1998adaptive} follow this approach have been widely used as benchmark problems in the \textsc{uqbo} literature (see \citeN{boros2007local} for an overview). In the sequel we discuss the impact of diagonal dominance, shifts, the density and extreme values of $\m F$ on the expected difficulty of the corresponding \textsc{uqbo} problem.

\subsubsection{Diagonal}
Generally, stronger \emph{diagonal dominance} in $\m F$ corresponds to easier \textsc{uqbo} problems \cite{billionnet1994minimization}. Consequently, the original problem generator presented by \citeN{pardalos1991construction} is designed to draw the off-diagonal elements from a uniform on a different support $\dset{-q,q}$ with $q\in\N$.

In this context, we point out that the impact of diagonal dominance carries over to the statistical properties of the tempered distributions \eqref{eq:tempered} we defined in the introductory Section \ref{sec:stat model}. Indeed, stronger diagonal dominance in $\m F$ corresponds to exponential quadratic distributions 
\begin{equation*}
\pi(\v\gamma)\eqdef\frac{\exp(\v\gamma\t\m F\v\gamma)}{\sum_{\v\gamma\in\B^d}\exp(\v\gamma\t\m F\v\gamma)}
\end{equation*}
having lower dependencies between the components of $\v\gamma$. We can analytically derive a parameter $\m A\in\R^{d\times d}$ for a logistic conditionals family $q^{\LogCo}_{\m A}$ that approximates $\pi(\v\gamma)$ where the quality of the approximation increases as the diagonal of $\m F$ becomes more dominant \cite{schaefer2012logistic}. We can accelerate the sequential Monte Carlo algorithm by initializing the system from $q^{\LogCo}_{\m A}$ instead of $\uni_{\B^d}$. However, we did not exploit this option to keep the present work more concise.

For positive definite $\m F\succ0$, the optimization problem is convex and can be solved in polynomial time \cite{kozlov1979polynomial}; in exact optimization, this fact is exploited to construct upper bounds for maximization problems \cite{poljak1995convex}. We observe a corresponding complexity reduction in statistical modeling. For $\m F\succ0$, the auxiliary distribution
\begin{equation*}
\pi(\v\gamma)\eqdef\frac{\v\gamma\t\m F\v\gamma}{2^{d-2}\left(\v 1\t \m F \v 1 + \tr{\m F}\right)},
\end{equation*}
is a feasible mass function, and we can derive analytical expressions concerning all cross-moments and marginal distributions \cite{schaefer2011parametric} which allows to largely analyze the properties of $\pi(\v\gamma)$ without enumerating the state space.

\subsubsection{Shifts}
The global optimum of the \textsc{uqbo} problem is more difficult to detect as we shift the entries of the matrix $\m F$ but the relative gap between the optimum and any heuristic value diminishes. If we sample $f_{ij}=f^{\tau}_{ij}$ from a uniform on the \emph{shifted} support
\begin{equation*}
q_{c,\tau}(k)\eqdef\uni_{\dset{-c+\tau,c+\tau}}(k), \quad c\in\N,\,\tau\in\dset{-c,c},
\end{equation*}
we obtain an objective function 
\begin{equation*}
f_{\tau}(\v x)
=\v x\t\m F^{\tau}\v x
\stackrel{d}{=}\v x\t(\m F^{0}+\tau\v1\v1\t)\v x
=f_{0}(\v x)+\tau\abs{\v x}^{2},
\end{equation*}
where $\stackrel{d}{=}$ means equality in distribution. Hence, with growing $\abs{\tau}$ the optimum depends less on $\m F$ and the relative gap between the optimum and a solution provided by any meta-heuristic vanishes. \citeN{boros2007local} define a related criterion
\begin{equation*}
\bar\rho\eqdef
\frac{1}{2}+\frac{\tau+2\tau c}{2(\tau^2+c^2+c)}\in[0,1]
\end{equation*}
and report a significant impact of $\bar\rho$ on the solution quality of their local search algorithms which is not surprising.

\subsubsection{Density}
The difficulty of the optimization problem is related to the number of interactions, that is the number of non-zero elements of $\m F$. We call the proportion of non-zeros the \emph{density} of $\m F$. Drawing $f_{ij}$ from the mixture
\begin{equation*}
q_{c,\omega}(k)=\omega\,\uni_{\dset{-c,c}}(k)+(1-\omega)\delta_{0}(k), \quad c\in\N,\,\omega\in(0,1]
\end{equation*}
we adjust the difficulty of the problem to a given expected density $\omega$.

Note that not all algorithms are equally sensitive to the density of $\m F$. Using the basic linearization \eqref{eq:lin ubqo}, each non-zero off-diagonal element requires the introduction of an auxiliary variable and three constraints. Thus, the expected total number of variables and the expected total number of constraints, which largely determine the complexity of the optimization problem, are proportional to the density $\omega$.

On the other hand, many randomized approaches, including the particle algorithms discussed in Section \ref{sec:smc}, are less sensitive to the density of the problem in the sense that replacing zero elements by small values has a minor impact on the performance of these algorithms. Rather than the zero/non-zero duality, we suggest that the presence of extreme values determines the difficulty of providing heuristic solutions.

\subsubsection{Extreme values}
\label{sec:extreme}
The uniform sampling approach advocated by \citeN{pardalos1991construction} is widely used in the literature for comparing meta-heuristics. Certainly, particle-driven methods are computationally too expensive to outperform local search heuristics on test problems with uniformly drawn entries; \cite{beasley1998heuristic} confirms this intuition with respect to genetic algorithms versus tabu search and simulated annealing. However, the uniform distribution does not produce \emph{extreme values} and it is vital to keep in mind that these have an enormous impact on the performance of local search algorithms.

Extreme values in $\m F$ lead to the existence of distinct local maxima $\v x^{*}\in\B^d$ of $f$ in the sense that there is no better candidate solution than $\v x^{*}$ in the neighborhood $\cup_{i=1}^{k}N_{i}(\v x^{*})$ even for relatively large $k$. Further, extreme local minima might completely prevent a local search heuristic from traversing the state space in certain directions. Consequently, local search algorithms, as discussed in Section \ref{sec:meta}, depend more heavily on their starting value, and their performance deteriorates with respect to particle-driven algorithms.

\begin{figure}[ht]
\caption{Histograms of a Cauchy $\cau_{5}$ and a uniform $\uni_{10}$ distribution.\vspace{-5mm}}
\label{fig:problem distr}
\begin{center}
\includegraphics[width=\textwidth]{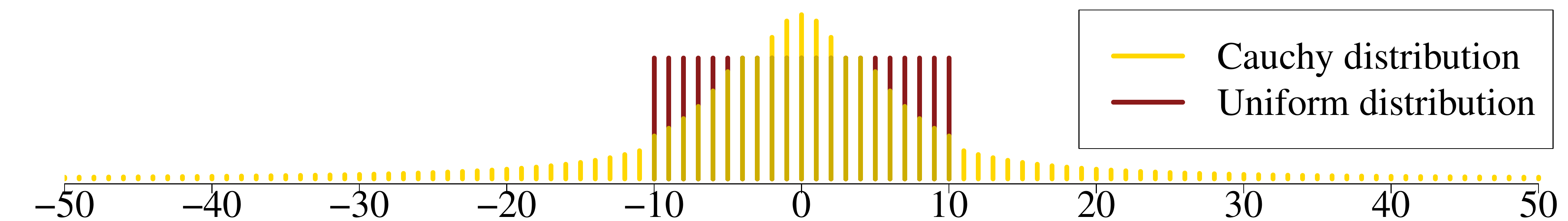}
\end{center}
\end{figure}

We propose to draw the matrix entries $f_{ij}$ from a discretized Cauchy distribution
\begin{equation}
\label{eq:cauchy}
\cau_{c}(k)\propto(1+(k/c)^2)^{-1}, \quad c\in\N
\end{equation}
that has heavy tails which cause extreme values to be frequently sampled. Figure \ref{fig:problem distr} shows the distribution of a Cauchy and a uniform to illustrate the difference. The resulting \textsc{uqbo} problems have quite distinct local maxima; in that case we also say that the function $f(\v x)$ is \emph{strongly multi-modal}.

\subsection{Numerical comparison}
\label{sec:outline num ex}
In this section, we provide numerical comparisons based on instances of the \textsc{uqbo} problem. We generated two random test suites of dimension $d=250$, each having $10$ instances. For the first suite, we sampled the matrix entries from a uniform distribution $\uni_{100}$ on $\dset{-100,100}$; for the second, we sampled from a Cauchy distribution $\cau_{100}$ as defined in \eqref{eq:cauchy}. For performance evaluation, we run a specified algorithm $100$ times on the same problem and denote the outcome by $\v x_1,\dots,\v x_{100}$.

\subsubsection{Visualization}
Since the absolute values are not meaningful, we report the relative ratios
\begin{equation*}
\varrho_k\eqdef\frac{f(\v x_k)-\text{worst solution found}}{\text{best known solution}-\text{worst solution found}}\in[0,1],
\end{equation*}
where the best known solution is the highest objective value ever found for that instance and the worst solution is the lowest objective value among the $100$ outcomes. We summarize the results in a histogram. The first $n$ bins are singletons $b_{k}\eqdef\set{\varrho_k^{*}}$ for the highest values $\varrho_1^{*}>\cdots>\varrho_{n}^{*}\in\set{\varrho_k\colon k\in\dset{1,100}}$; the following $n$ bins are equidistant intervals $b_k^{<}\eqdef[\frac{n-k}{n}\varrho_{n}^{*},\frac{n-k+1}{n}\varrho_{n}^{*})$. The graphs show the bins $b_{1},\dots,b_{n},b_{1}^{<},\dots,b_{n}^{<}$ in descending order from left to right on the $x$-axis. The interval bins are marked with a sign ``$<$'' and the lower bound. The $y$-axis represents the counts.

For comparison, we draw the outcome of several algorithms into the same histogram, where the worst solution found is the lowest overall objective value among the outcomes. For each algorithm, the counts are depicted in a different color and, for better readability, with diagonal stripes in a different angle. To put it plainly, an algorithm performs well if its boxes are on the left of the graph since this implies that the outcomes where often close to the best known solution.

\subsubsection{Comparison of binary parametric families}
\label{sec:perf fam}
We study how the choice of the binary parametric family affects the quality of the delivered solutions. The focus is on the cross-entropy method, since we cannot easily use the Gaussian copula family in the context of sequential Monte Carlo. We use $n=1.2\times10^4$ particles, set the speed parameter to $\beta=0.8$ (or the elite fraction to $0.2$) and the lag parameter to $\tau=0.5$.

\begin{figure*}[ht]
\begin{centering}
\caption{The cross-entropy method using different binary parametric families.}
\subfigure[problem $f(\v x)=\v x\t\m F\v x$ with $f_{ij}\sim\cau_{100}$ for $i,j\in\dset{1,250}$]{
\includegraphics[width=0.95\textwidth]{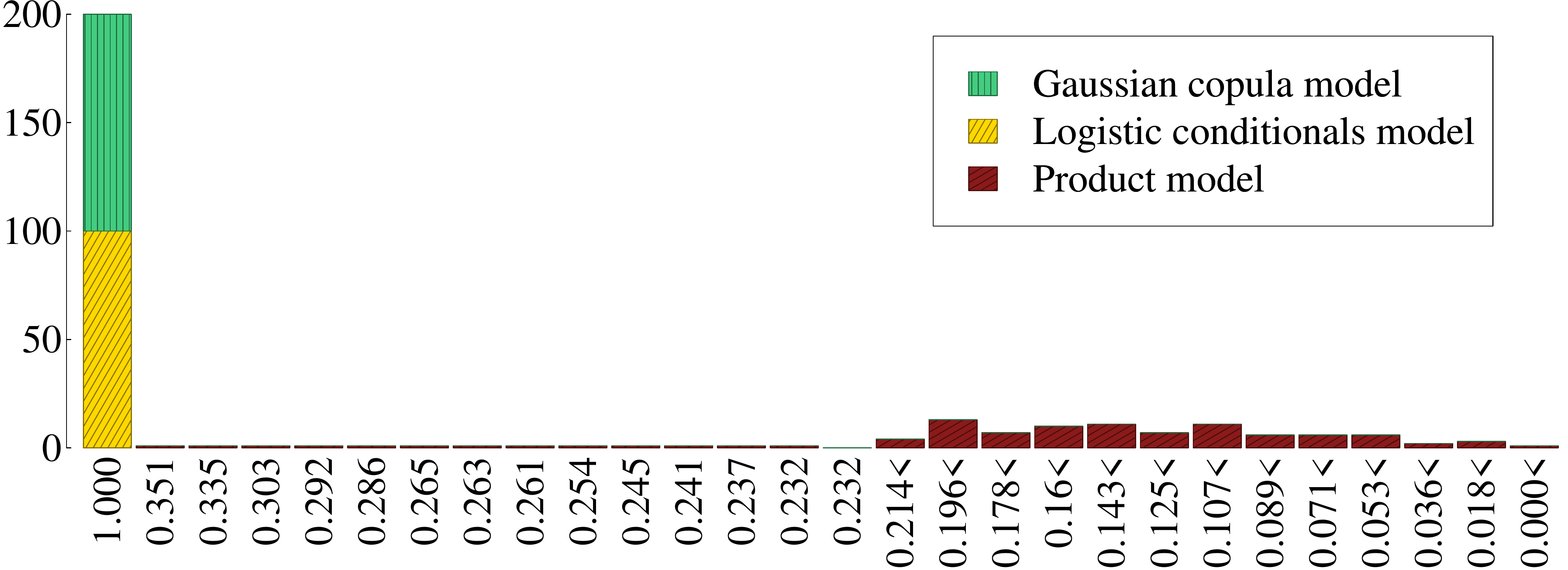}
\label{fig:cauchy mc}
}
\subfigure[problem $f(\v x)=\v x\t\m F\v x$ with $f_{ij}\sim\uni_{100}$ for $i,j\in\dset{1,250}$]{
\includegraphics[width=0.95\textwidth]{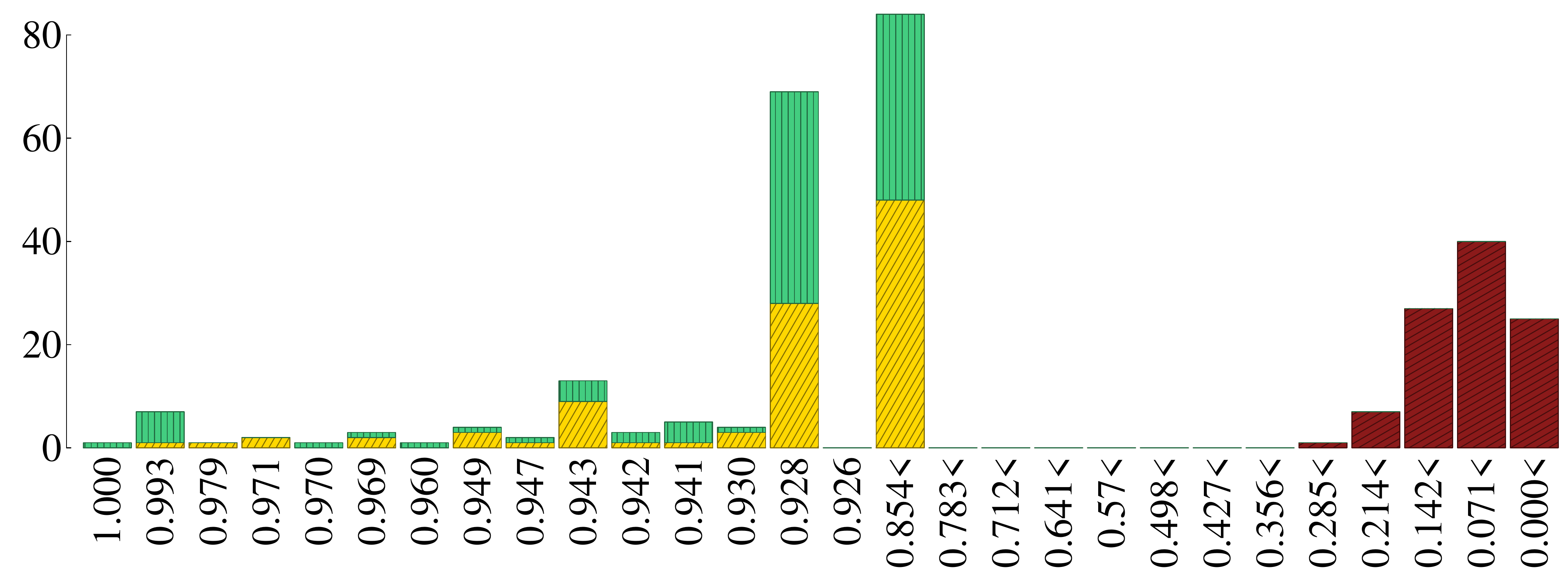}
\label{fig:uni mc}
}
\end{centering}
\end{figure*}

The numerical comparisons, given in Figures \ref{fig:uni mc} and \ref{fig:cauchy mc}, clearly suggest that using more advanced binary parametric families allows the cross-entropy method to detect local maxima that are superior to those detected using the product family. Hence, the numerical experiments confirm the intuition of our toy example in Figure \ref{fig:toy exa}.

On the strongly multi-modal instance \ref{fig:cauchy mc} the numerical evidence for this conjecture is stunningly clear-cut; on the weakly multi-modal problem \ref{fig:uni mc} its validity is still unquestionable. This result seems natural since reproducing the dependencies induced by the objective function is more relevant in the former case than in the latter.

\subsubsection{Comparison of optimization algorithms}
\label{sec:perf algo}
We compare a sequential Monte Carlo sampler with parametric family, a sequential Monte Carlo sampler with a single-flip symmetric kernel \eqref{eq:sym kernel}, the cross-entropy method, simulated annealing and $1$-opt local search as described in Section \ref{sec:algorithms}.

For the cross entropy method, we use the same parameters as in the preceding section. For the sequential Monte Carlo algorithm, we use $n=0.8\times10^4$ particles and set the speed parameter to $\beta=0.9$; we target a tempered auxiliary sequence \eqref{eq:tempered}. For both algorithms we use the logistic conditionals family as sampling distribution. With these configurations, the algorithms converge in roughly $25$ minutes. We calibrate the sequential Monte Carlo sampler with local moves to have the same average run time by processing batches of $10$ local moves before checking the particle diversity criterion. The simulated annealing and $1$-opt local search algorithms run for exactly $25$ minutes.

\begin{figure*}[ht]
\begin{centering}
\caption{Comparison of stochastic optimization algorithms on two \textsc{uqbo} problems.}
\subfigure[problem $f(\v x)=\v x\t\m F\v x$ with $f_{ij}\sim\cau_{100}$  for $i,j\in\dset{1,250}$]{
\includegraphics[width=0.95\textwidth]{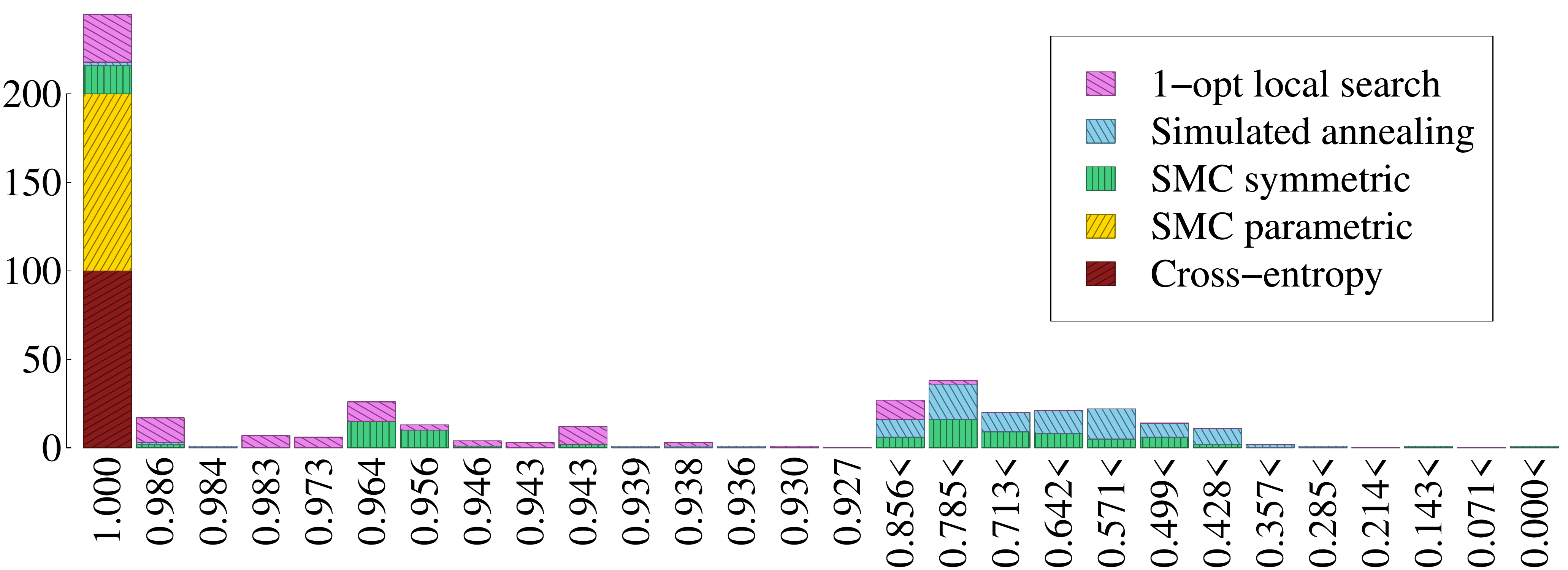}
\label{fig:cauchy ac}
}
\subfigure[problem $f(\v x)=\v x\t\m F\v x$ with $f_{ij}\sim\uni_{100}$  for $i,j\in\dset{1,250}$]{
\includegraphics[width=0.95\textwidth]{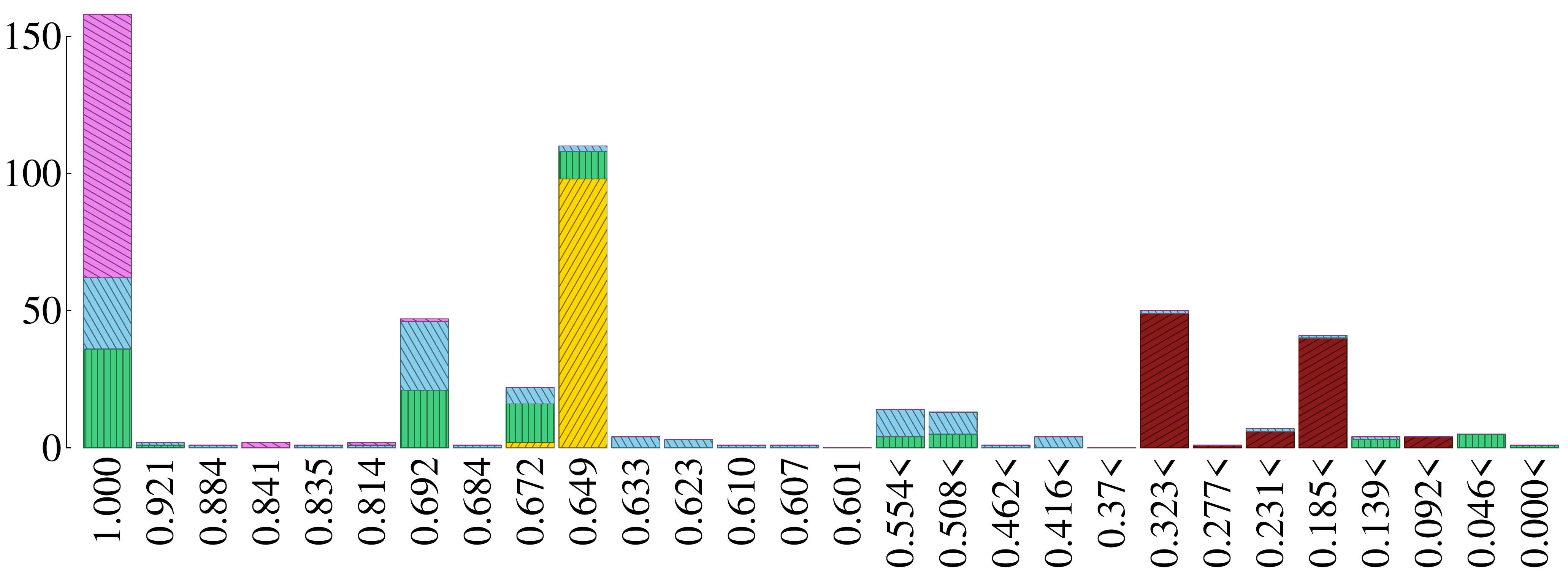} 
\label{fig:uni ac}
}
\end{centering}
\end{figure*}

The results shown in Figures \ref{fig:uni ac} and \ref{fig:cauchy ac} assert the intuition that particle methods perform significantly better on strongly multi-modal problems. However, on the easy test problems, the particle methods tend to persistently converge to the same sub-optimal local modes. This effect is probably due to their poor local exploration properties. Since particle methods perform significantly less evaluations of the objective function, they are less likely to discover the highest peak in a region of rather flat local modes. The use of parametric families aggravates this effect, and it seems advisable to alternate global and local moves to make a particle algorithm more robust against this kind of behavior.

Further numerical results are shown in Figure \ref{fig:r250c} and Figure \ref{fig:r250u}.


\begin{figure*}
\caption{Comparison of stochastic optimization algorithms. $10$ problems $f(\v x)=\v x\t\m F\v x$ with $f_{ij}\sim\cau_{100}$ for $i,j\in\dset{1,250}$}
\label{fig:r250c}
\mygraph{r250c}{01}
\mygraph{r250c}{02}
\mygraph{r250c}{03}
\mygraph{r250c}{04}
\mygraph{r250c}{05}
\mygraph{r250c}{06}
\mygraph{r250c}{07}
\mygraph{r250c}{08}
\mygraph{r250c}{09}
\mygraph{r250c}{10}
\end{figure*}

\begin{figure*}
\caption{Comparison of stochastic optimization algorithms. $10$ problems $f(\v x)=\v x\t\m F\v x$ with $f_{ij}\sim\uni_{100}$ for $i,j\in\dset{1,250}$}
\label{fig:r250u}
\mygraph{r250u}{01}
\mygraph{r250u}{02}
\mygraph{r250u}{03}
\mygraph{r250u}{04}
\mygraph{r250u}{05}
\mygraph{r250u}{06}
\mygraph{r250u}{07}
\mygraph{r250u}{08}
\mygraph{r250u}{09}
\mygraph{r250u}{10}
\end{figure*}


\section{Discussion and conclusion}
The numerical experiments carried out on different parametric families revealed that the use of the advanced families proposed in this paper significantly improves the performance of the particle algorithms, especially on the strongly multi-modal problems. The experiments demonstrate that local search algorithms, like simulated annealing and randomized $1$-opt local search, indeed outperform particle methods on weakly multi-modal problems but deliver inferior results on strongly multi-modal problems.

Using tabu lists, adaptive restarts and rounding heuristics, we can certainly design local search algorithms that perform better than simulated annealing and $1$-opt local search. Still, the structural problem of strong multi-modality persists for path-based algorithms. On the other hand, cleverly designed local search heuristics will clearly beat sequential Monte Carlo methods on easy to moderately difficult problems.

The results encourage the use of particle methods if the objective function is known to be potentially multi-modal and hard to analyze analytically. We have to keep in mind that multiple restarts of rather simple local search heuristics can be very efficient if they make use of the structure of the objective function. For $25$ minutes of randomized restarts, the heuristic proposed by \citeN{boros2007local}, which exploits the fact that the partial derivatives of a multi-linear function are constant, practically always returns the best known solution on all test problems treated to create Figures \ref{fig:r250c} and \ref{fig:r250u}.

The numerical work was completely done in \href{http://www.python.org/}{Python 2.6} using \href{http://www.scipy.org}{SciPy} packages and run on a cluster with $1.86$ GHz processors. The sources used in this work and the problems processed in this paper can be found at \url{http://code.google.com/p/smcdss}.

\section{Acknowledgments}
This work is part of the author's Ph.D. thesis at CREST under supervision of Nicolas Chopin whom I would like to thank for the numerous discussions on particle algorithms. I thank the editor and two anonymous referees for their detailed comments which helped to significantly improve this paper.

\section{Appendix: Fitting the parameters}
We briefly summarize how the parameters of the logistic conditionals family and the Gaussian copula family can be assessed for a given particle system $\m X=(\v x_1,\dots,\v x_n)\t$ with weights $\v w=(w_{1},\dots,w_{n})$. We denote by
\begin{equation}
\label{eq:moments}
\textstyle\bar x_i\eqdef\sum_{k=1}^n w_k x_{ki},\quad \bar x_{ij}\eqdef\sum_{k=1}^n w_k x_{ki}x_{kj},\quad i,j\in \dset{1,d}
\end{equation}
the weighted first and second sample moments.

\subsection{Logistic conditionals family}
\subsubsection{Derivatives of the log-likelihood function}
The log-likelihood function of the weighted logistic regression of $\v y^{(i)}\eqdef \m X_{\bullet i}$ on $\m Z^{(i)}\eqdef(\m X_{\bullet1:i-1},\v 1)$ is
\begin{align*}
\log L(\v a)
&=\sum_{k=1}^n w_k \left[y^{(i)}_k\log[\logistic(\v z_{k \bullet}^{(i)}\v a)]+(1-y^{(i)}_k)\log[1-\logistic(\v z_{k \bullet}^{(i)}\v a)]\right] \\
&=\sum_{k=1}^n w_k\left[y^{(i)}_k\v z_{k \bullet}^{(i)}\v a-\log[1+\exp(\v z_{k \bullet}^{(i)}\v a)]\right],
\end{align*}
where we used that $\log[1-\logistic(\v x\t\v a)]=-\log[1+\exp(\v x\t\v a)]=-\v x\t\v a+\log[\logistic(\v x\t\v a)]$. Since $\partial\log[1+\exp(\v x\t \v a)]/\partial\v a=\logistic(\v x\t\v a)\v x$, the gradient of the log-likelihood is
\begin{align*}
s(\v a)
=\sum_{k=1}^n w_k\left[y^{(i)}_k\v z_{k \bullet}^{(i)}-\logistic(\v z_{k \bullet}^{(i)}\v a)\v z_{k \bullet}^{(i)}\right]
=(\m Z^{(i)})\t \diag(\v w)[\v y^{(i)} - \v p^{(i)}_{\v a}],
\end{align*}
where $(p^{(i)}_{\v a})_k\eqdef\logistic(\v z_{k \bullet}^{(i)}\v a)$. Since $\partial\logistic(\v x\t \v a)/\partial\v a=-\logistic(\v x\t \v a)[1-\logistic(\v x\t \v a)]\v x$, the Hessian matrix of the log-likelihood is
\begin{align*}
s'(\v a)
=-\sum_{k=1}^n w_k\left[\logistic(\v z_{k \bullet}^{(i)}\v a)[1-\logistic(\v z_{k \bullet}^{(i)}\v a)]\right]\v z_{k \bullet}^{(i)}(\v z_{k \bullet}^{(i)})\t
=-(\m Z^{(i)})\t \diag(\v w) \diag(\v q^{(i)}_{\v a}) \m Z^{(i)},
\end{align*}
where $(q^{(i)}_{\v a})_k\eqdef\logistic(\v z_{k \bullet}^{(i)}\v a)[1-\logistic(\v z_{k \bullet}^{(i)}\v a)]$.

\subsubsection{Complete separation}
The data might suffer from complete or quasi-complete separation \cite{albert_84} which causes the likelihood function $L(\v a)$ to be monotonic. In that case there is no maximizer. We can avoid the monotonicity by assigning a suitable prior distribution to the parameter $\v a$. \citeN{firth_93} recommends the Jeffreys prior for its bias reduction which can conveniently be implemented via data adjustment \cite{kosmidis2009bias}.

For the sake of simplicity, however, we only assign a simple Gaussian prior with variance $\varepsilon^{-1}>0$ such that, up to a constant, the log-posterior distribution is the log-likelihood function plus a quadratic penalty term and therefore always convex. The score function and its Jacobian matrix become
\begin{align*}
s(\v a) =(\m Z^{(i)})\t \diag(\v w)[\v y^{(i)} - \v p^{(i)}_{\v a}] - \varepsilon\v a,\quad
s'(\v a)=-(\m Z^{(i)})\t \diag(\v w) \diag(\v q^{(i)}_{\v a}) \m Z^{(i)}-\varepsilon \m I.
\end{align*}
The bias-reduced estimators are known to shrink towards $0.5$ which is an undesired property when fitting a parametric family. Therefore, we attempt to keep the shrinkage parameter $\varepsilon$ as small as possible.

\subsubsection{Newton-Raphson iterations}
The first order condition $s(\v a)=\v 0$ is solved iteratively
\begin{align*}
\v a^{(t+1)}
=\v a^{(t)}-[s'(\v a^{(t)})]^{-1}s(\v a^{(t)})
=\v a^{(t)}+\v x^{(t)}
\end{align*}
where $\v x^{(t)}$ is the vector that solves
\begin{align*}
\left[(\m Z^{(i)})\t \diag(\v w) \diag(\v q^{(i)}_{\v a^{(t)}})(\m Z^{(i)})\t+\varepsilon\m I\right]\v x^{(t)}=
\left[(\m Z^{(i)})\t \diag(\v w) [\v y^{(i)} - \v p^{(i)}_{\v a^{(t)}}]-\varepsilon\v a^{(t)}\right].\end{align*}
If the Newton iteration at the $i$th component fails to converge, we can either augment the penalty term $\varepsilon$ which leads to stronger shrinkage of the mean $m_i$ towards $0.5$ or we can drop some covariates $\gamma_j$ for $j\in\set{1,\dots,i-1}$ from the iteration to improve the numerical condition of the procedure.

In practice, we drop the predictors from the regression model which are only weakly correlated with the explained variable. This step is important to speed up the algorithm and improve its numerical properties. For a proposal distribution, it is particularly important to take the strong dependencies into account but it is often sufficient to work with very sparse logistic conditionals families.

In particularly difficult cases, we might prefer to set $a_{ii}=\logistic^{-1}(\bar x_{i})$ and $\v a_{i,1:i-1}=\v 0$, where $\bar x_{i}$ is defined in \eqref{eq:moments}, which guarantees that at least the mean is correct. This is an important issue since misspecification of the mean of $\gamma_i$ also affects the distribution of the components $\gamma_j$ for $j\in\set{i+1,\dots,d}$ which are sampled conditional on $\gamma_i$.


\begin{algorithm}
\DontPrintSemicolon
\KwIn{$\m X=(\v x_1,\dots,\v x_n)\t,\ \v w=(w_1,\dots,w_n),\ \m A\in\R^{d\times d}$}\vspace{0.3em}
\For {$i\in\dset{1,d}$}{$ $\\[0.3em]
   $\m Z\gets(\m X_{\bullet1:i-1},\v 1),\ \v y\gets\m X_{\bullet i}, \ \v a^{(0)}\gets\m A_{i,1:i}$\\
  \Repeat{$\Vert\v a^{(t+1)}-\v a^{(t)}\Vert_{\infty}<\delta$}{
    \begin{tabular}{rllr}
    $p_k$       &\hspace{-3mm}$\gets$&\hspace{-3mm}$\logistic(\m Z_{k\bullet} \v a^{(t)})$&\textbf{ for all }$k\in\dset{1,n}$ \\
    $q_k$       &\hspace{-3mm}$\gets$&\hspace{-3mm}$p_k(1-p_k)$                    &\textbf{ for all }$k\in\dset{1,n}$ \\
    \end{tabular}
    \begin{tabular}{ll}
    $\v a^{(t+1)}\gets$&\hspace{-3mm}$\v a^{(t)}+\left[(\m Z^{(i)})\t \diag[\v w] \diag[\v q]\m Z^{(i)}+\varepsilon\m I\right]^{-1}\times$\\
    &$\qquad\qquad\left[(\m Z^{(i)})\t \diag[\v w] \left[\v y - \v p \right]-\varepsilon\v a^{(t)}\right]$\\
    \end{tabular}
  }\vspace{0.3em}
  $\m A_{i,1:i}\gets\v a$ \\
}
\Return $\m A$
\caption{Fitting the weighted logistic regressions}
\label{algo:fit logistic}
\end{algorithm}

\subsection{Gaussian copula family}
We adjust the Gaussian copula family $q^{\Gau}_{\v a,\m \Sigma}$ by method of moments. We need to solve the non-linear equations
\begin{equation}
\label{eq:fitting Gaussian}
\varPhi_1(a_i)=\bar x_i,\quad \varPhi_2(a_i,a_j;\sigma_{ij})=\bar x_{ij},\quad i,j\in\dset{1,d},
\end{equation}
where $\bar x_i$ and $\bar x_{ij}$ are defined in \eqref{eq:moments} while $\varPhi_1(v_i)$ and $\varPhi_2(v_i,v_j;\sigma_{ij})$ denote the cumulative distribution functions of the univariate and bivariate normal distributions with zero mean, unit variance and correlation coefficient $\sigma_{ij}\in[-1,1]$.

While the parameter $a_i=\varPhi_1^{-1}(\bar x_i)$ is easy to assess, the challenging task is to compute the bivariate variances $\sigma_{ij}$ for all $i,j\in \dset{1,d}$. Recall the standard result [\citeNP{johnson2002continuous}, p.255]
\begin{equation}
\label{eq:der of norm pdf}
\frac{\partial \varPhi_2(x_1,x_2;\sigma)}{\partial\sigma}=\varphi(x_1,x_2;\sigma),
\end{equation}
where $\varphi(\cdot,\cdot; \sigma)$ denotes the density of the bivariate normal distribution with correlation coefficient $\sigma$. We obtain the following Newton-Raphson iteration
\begin{equation}
\sigma^{(t+1)}=\sigma^{(t)}-
\frac{\varPhi_2(a_i,a_j;\sigma_{ij}^{(t)})-\bar x_{ij}}{\varphi(a_i,a_j; \sigma_{ij}^{(t)})},
\end{equation}
starting at some initial value $\sigma_{ij}^{(0)}\in(-1, 1)$; see Procedure \ref{algo:fit gaussian}. In the sequential Monte Carlo context, good initial values are obtained from the parameters of the previous auxiliary distributions.

We use a fast series approximation \cite{drezner_98} to evaluate $\varPhi_2(a_i,a_j;\sigma_{ij})$. These approximations are critical when $\sigma_{ij}$ comes very close to either boundary of $[-1,1]$. The Newton iteration might repeatedly fail when restarted at the corresponding boundary $\sigma_{ij}^{(0)}\in\set{-1,1}$. In any event, $\varPhi_2(x_1,x_2;\sigma)$ is strictly monotonic in $\sigma$ since its derivative \eqref{eq:der of norm pdf} is positive, and we can switch to bi-sectional search if necessary.

\begin{algorithm}
\DontPrintSemicolon
\KwIn{$\bar x_i,\ \bar x_{ij}\ $\textbf{ for all }$i,j\in \dset{1,d}$}\vspace{0.3em}
$a_i\gets\varPhi^{-1}(\bar x_i)\ $\textbf{ for all }$i\in \dset{1,d}$\\[0.3em]
$\m \Sigma^{(0)}\gets\m I$ \\[0.3em]
\For {$i,j\in \dset{1,d}, \ i<j$}{\vspace{0.3em}
  \Repeat{$\vert \sigma_{ij}^{(t+1)}-\sigma_{ij}^{(t)}\vert<\delta$}{
   $\displaystyle \sigma_{ij}^{(t+1)}\gets
      \sigma_{ij}^{(t)}-\frac{\varPhi_2(a_i,a_j;\sigma_{ij}^{(t)})-\bar x_{ij}}{\varphi(a_i,a_j; \sigma_{ij}^{(t)})}$ \\
  }\vspace{0.3em}
  $\sigma_{ji}\gets\sigma_{ij}^{(t+1)}$ \\[0.3em]
}
\textbf{if not }$\m\Sigma\succ 0$\textbf{ then }$\m \Sigma\gets(\m \Sigma+\abs{\lambda}\m I)/(1+\abs{\lambda})$ \\
\Return $\v a,\, \m \Sigma$
\caption{Fitting the dependency matrix}
\label{algo:fit gaussian}
\end{algorithm}

The locally fitted correlation matrices $\m \Sigma$ might not be positive definite for $d \geq 3$ because the Gaussian copula is not flexible enough to model the full range of cross-moments binary distributions might have (see \citeN{schaefer2012logistic} for an extended discussion). We obtain a feasible parameter replacing $\m \Sigma$ by
\begin{equation*}
\m \Sigma^*=(\m \Sigma+\abs{\lambda}\m I)/(1+\abs{\lambda}), 
\end{equation*}
where $\lambda$ is smaller than all eigenvalues of the locally fitted matrix $\m \Sigma$. This approach evenly lowers the local correlations to a feasible level and is easy to implement on standard software. In practice, we also prefer to work with a sparse version of the Gaussian copula family that concentrates on strong dependencies and sets minor correlations to zero.


\end{document}